\documentclass[
superscriptaddress,
twocolumn,
amsmath,amssymb,
aps,
pra,
]{revtex4-2}

\usepackage{graphicx,framed} 
\usepackage{dcolumn} 
\usepackage{bm} 
\usepackage{braket}
\usepackage{color}
\usepackage{amsthm}
\usepackage{hyperref} 
\usepackage[normalem]{ulem} 
\usepackage{qcircuit}       

\usepackage{amsthm}

\usepackage{comment}

\usepackage{float}
\makeatletter
\let\newfloat\newfloat@ltx
\makeatother
\usepackage{algorithm}
\usepackage[noend]{algpseudocode}
\algdef{SE}[DOWHILE]{Do}{doWhile}{\algorithmicdo}[1]{\algorithmicwhile\ #1}
\makeatletter
\def\algbackskip{\hskip-\ALG@thistlm}
\makeatother

\definecolor{lightblue}{RGB}{73,151,208}
\definecolor{crimson}{RGB}{140,41,53}

\newcommand{\hl}{\color{black}}

\hypersetup{
    colorlinks,
    linkcolor={crimson},
    citecolor={lightblue},
    urlcolor={lightblue}
}

\newcommand{\dt}{\delta{t}}

\begin{document}

\preprint{}

\title{{\hl Measuring Trotter error and its application to precision-guaranteed Hamiltonian simulations}}

\author{Tatsuhiko N. Ikeda}
\email{tatsuhiko.ikeda@riken.jp}
\affiliation{RIKEN Center for Quantum Computing, Wako, Saitama 351-0198, Japan}
\affiliation{Department of Physics, Boston University, Boston, Massachusetts 02215, USA}

\author{Hideki Kono}
\email{konofr0924@g.ecc.u-tokyo.ac.jp}
\affiliation{Department of Applied Physics, The University of Tokyo, Hongo, Tokyo, 113-8656, Japan}
\affiliation{RIKEN Center for Quantum Computing, Wako, Saitama 351-0198, Japan}

\author{Keisuke Fujii}
\email{keisuke.fujii.ay@riken.jp}
\affiliation{Graduate School of Engineering Science, Osaka University,
1-3 Machikaneyama, Toyonaka, Osaka 560-8531, Japan.}
\affiliation{Center for Quantum Information and Quantum Biology, Osaka University, 560-0043, Japan.}
\affiliation{RIKEN Center for Quantum Computing, Wako, Saitama 351-0198, Japan}
\affiliation{Fujitsu Quantum Computing Joint Research Division at QIQB,
Osaka University, 1-2 Machikaneyama, Toyonaka 560-0043, Japan}
\date{\today}%

\begin{abstract}
{\hl
Trotterization is the most common and convenient approximation method for Hamiltonian simulations on digital quantum computers, but estimating its error accurately is computationally difficult for large quantum systems. Here, we develop a method for measuring the Trotter error without ancillary qubits on quantum circuits by combining the $m$th- and $n$th-order ($m<n$) Trotterizations rather than consulting with mathematical error bounds. Using this method, we make Trotterization precision-guaranteed, developing an algorithm named Trotter$(m,n)$, in which the Trotter error at each time step is within an error tolerance $\epsilon$ preset for our purpose. Trotter$(m,n)$ is applicable to both time-independent and -dependent Hamiltonians, and it adaptively chooses almost the largest stepsize $\dt$, which keeps quantum circuits shallowest, within the error tolerance. Benchmarking it in a quantum spin chain, we find the adaptively chosen $\dt$ to be about ten times larger than that inferred from known upper bounds of Trotter errors.
}
\end{abstract}

\maketitle

\section{Introduction}
The rapid development of quantum devices in recent years has led researchers to find useful applications with significant quantum advantage~\cite{Harrow2017,Cao2019,Cerezo2021}.
On top of the eigenvalue problems~\cite{Aspuru-Guzik2005,Lloyd1996,Whitfield2011,Higgott2019,Jones2019,Kirby2021},
quantum many-body dynamics, or Hamiltonian simulation~\cite{Babbush2015,Low2017,Campbell2019,An2021,An2022,Childs2022}, is one of the most promising candidates because quantum computers could overcome the exponential complexity that classical computers face~\cite{Lloyd1996}, enabling us to address intriguing dynamical phenomena like nonequilibrium phases of matter~\cite{Swingle2016,Zhang2017,Landsman2019,Joshi2022} and to implement fundamental quantum algorithms like phase estimation~\cite{Kitaev1995}.
Among several algorithms for the Hamiltonian simulation, Trotterization~\cite{Trotter1959,Hatano2005} is and will be used most commonly in the current noisy intermediate-scale quantum (NISQ~\cite{Preskill2018}) era and the coming early fault-tolerant quantum computing (FTQC) era because it does not demand additional ancillary qubits or largely controlled quantum gates.
Indeed, quantum advantage in Trotterized dynamics simulation has been reported using a 127-qubit NISQ computer only recently~\cite{Kim2023}.

One major and presumably inevitable issue of Trotterization is the trade-off relation between the simulation accuracy and the circuit depth {\hl (see, however, Refs.~\cite{Barison2021,Mansuroglu2023} for variational approaches)}.
The $k$-th order Trotterization accompanies an error of $O(\dt^{k+1})$ during a single time step $\dt$, which decreases when $\dt$ is taken shorter.
In the meantime, the number of steps to reach a final time increases, meaning a deeper quantum circuit.
To suppress the gate depth, it is desirable to choose the largest possible stepsize $\dt$, i.e., the shallowest circuit, within our error tolerance $\epsilon$ preset for our purposes.

However, it is difficult to find the optimal stepsize $\dt$ because the Trotter error is complex in generic many-body systems.
According to the previous studies on the Trotter error, its upper bounds~\cite{Kivlichan2020,Childs2021} and typical values~\cite{Zhao2021} are available.
If we choose $\dt$ so that the upper bound is below our tolerance $\epsilon$, the precision is guaranteed, but $\dt$ tends to be too small, as we will see below.
On the other hand, if we choose $\dt$ based on the typical values, $\dt$ can be larger, but the precision guarantee is lost.
Recently, Zhao et al.~\cite{Zhao2022} proposed {\hl an approach} where $\delta t$ is chosen adaptively in each time step based on the energy expectation value and variance.
Yet, the precision guarantee of this method is still elusive, and the applicability is limited to time-independent Hamiltonians~\footnote{After the submission of our work, Zhao et al.~\cite{Zhao2023} publicized their adaptive-stepsize Trotterization generalized to time-dependent Hamiltonians.}.

\begin{figure*}[t]
    \centering
    \includegraphics[width=17cm]{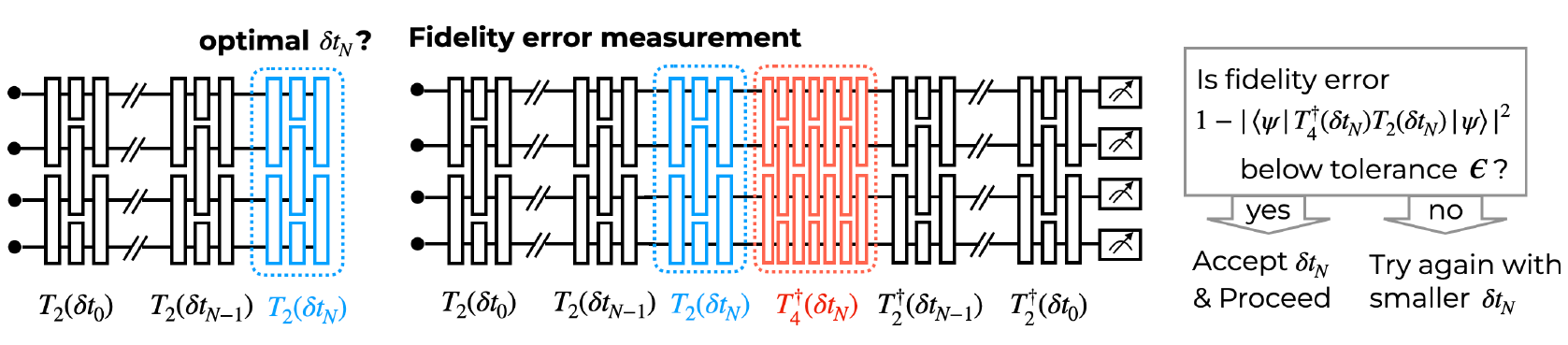}
    \caption{Key concept of (fidelity-based) Trotter24.
    (Left) At each time step $N$, we ask what the optimal stepsize $\delta t_N$ for the second-order Trotterization formula $T_2(\delta t_N)$. Here the optimal means the largest with the fidelity error kept less than our tolerance $\epsilon$.
    (Middle) The fidelity error that $T_2(\delta t_N)$ can be measured using the fourth-order Trotterization $T_4(\delta t_N)$, instead of the exact evolution $U(\delta t_N)$ when we neglect higher-order corrections in terms of $\delta t_N$.
    (Right) If the measured error is below our tolerance, we accept the value of $\delta t_N$ and proceed to the next time step $N+1$. Otherwise, we reject the value and try the same protocol with a smaller value for $\delta t_N$ again.
    In the paper, we also develop an observable-based, rather than fidelity-based, algorithm, establish an efficient scheme for avoiding the possible rejections of $\delta t_N$'s, and analyze how errors propagate with time steps.
    }
    \label{fig:outline}
\end{figure*}

{\hl In this paper, we develop a method to measure the Trotter error on quantum circuits by combining Trotterization formulas at different orders, $m$ and $n(>m)$. Since measured, the estimated error is significantly more accurate than known upper bounds for it and thus allows us to accurately choose the largest possible stepsize $\dt$ so that the error does not exceed our tolerance $\epsilon$.
Using this method, we make Trotterization precision-guaranteed, in which almost the largest $\dt$ is adaptively chosen within a preset error tolerance $\epsilon$ (see Fig.~\ref{fig:outline} for its structure).
We name this algorithm Trotter$(m,n)$ in analogy to RFK45 (Runge-Kutta-Fehlberg) for classical simulations, where the fourth- and fifth-order methods are combined.
}
We benchmark Trotter24 in a quantum spin chain under time-independent and -dependent Hamiltonians, finding that the adaptively chosen $\dt$ is about ten times larger than that inferred from the upper bound of Trotter errors.

\section{Measuring Trotter error}\label{sec:1step}
{\hl In this section, we present a way to measure a Trotter error on quantum circuits without ancillary qubits when a quantum state is evolved by an $m$-th order Trotterization for a small time step $\dt$. A key idea is making use of a higher-order $n(>m)$-th Trotterization to approximate the unknown exact solution accurately enough.}

\subsection{Extracting Trotter error using different orders}\label{sec:eta_eqs}
For simplicity, we first consider a time-independent Hamiltonian $H$ consisting of two parts,
\begin{align}\label{eq:AandB}
    H=A+B,
\end{align}
where $A$ and $B$ do not necessarily commute with each other.
Generalization to more noncommuting parts is straightforward, and we will generalize the arguments to time-dependent ones later in Sec.~\ref{sec:tdep}.
We assume that the quantum state at time $t$ is known to be $\ket{\psi(t)}$ and consider evolving it by a small time step $\delta t$,
\begin{align}
    \ket{\psi(t+\delta t)} = U(\dt)\ket{\psi(t)} = e^{-i H \delta t}\ket{\psi(t)}.
\end{align}
Trotterization approximately decomposes $e^{-i H\dt}$ into quantum-gate-friendly parts consisting of either $A$ and $B$.
{\hl We take an $m$-th order Trotterization $T_m(\dt)$.
For example, $T_m(\dt)$ can be the Lie-Trotter formula $T_1(\dt)=e^{-i A \dt}e^{-i B\dt}$ for $m=1$ or its symmetrized form $T_2(\delta t) \equiv e^{-i A \delta t/2}e^{-i B\delta t}e^{-i A \delta t/2}$ for $m=2$.
In general, $T_m(\dt)$ approximates $U(\dt)$ up to the order of $\dt^{m}$ and satisfies
\begin{align}
T_m(\delta t) = e^{-i H \delta t+\Upsilon_{m+1}}
\end{align}
where $\Upsilon_{m+1}=O(\delta t^{m+1})$ is an anti-Hermitian error operator.
These relations imply
\begin{align}\label{eq:def_psi2}
    \ket{\psi_m(t+\delta t)} \equiv T_m(\delta t)\ket{\psi(t)}
    = \ket{\psi(t+\delta t)} +O(\dt^{m+1}),
\end{align}
meaning that $T_m(\dt)$ approximates the exact one-step evolution within an error of $O(\dt^{m+1})$.
}

To quantify the error arising in the one step, we adopt the {\hl square root of the infidelity}
\begin{align}\label{eq:def_error_F}
{\hl    \eta_F \equiv \sqrt{1-|\braket{\psi(t+\delta t)|\psi_m(t+\delta t)}|^2}.}
\end{align}
We can also use other quantities depending on our purposes and make parallel arguments.
For example, when we are interested in the expectation value of an observable $O$, we care about the error in it,
\begin{align}
    \eta_O \equiv \braket{\psi(t+\delta t)|O|\psi(t+\delta t)}-\braket{\psi_m(t+\delta t)|O|\psi_m(t+\delta t)}.
\end{align}
In either case, calculating $\eta_F$ or $\eta_O$ is difficult because we do not know the exactly evolved state $\ket{\psi(t+\dt)}$.

We remark that {\hl both $\eta_F$ and $\eta_O$ are $O(\dt^{m+1})$, but showing $\eta_F=O(\dt^{m+1})$ is less obvious.}
To show this, we note that the leading $O(\dt^{m+1})$ term of $1-\braket{\psi(t+\delta t)|\psi_m(t+\delta t)}$ is pure-imaginary as shown in Appendix~\ref{app:error},
and its leading-order contribution of $\eta_F$ is given by
{\hl 
\begin{align}\label{eq:eta_F_leading}
\eta_F&=\sqrt{\braket{\psi(t)|(i\Upsilon_{m+1})^2|\psi(t)}-\braket{\psi(t)|(i\Upsilon_{m+1})|\psi(t)}^2} \notag\\
&\qquad+O(\dt^{m+2}),
\end{align}
where $i\Upsilon_{m+1}$ is Hermitian.}
Equation~\eqref{eq:eta_F_leading} dictates that $\eta_F$ is the variance of the ``observable'' $i\Upsilon_{m+1}$,
giving a way to estimate $\eta_F$ using $\ket{\psi(t)}$ and the explicit form of $i\Upsilon_{m+1}$.
Indeed this is a possible way of measuring $\eta_F$, but it requires, for generic many-body Hamiltonians, measuring numerous Hermitian operators involved in $i\Upsilon_{m+1}$ consisting of doubly nested commutators between $A$ and $B$.
{\hl Hence, in the following subsection, we will also discuss another way to estimate $\eta_F$ with less sampling costs.}

{\hl
Our idea of estimating the errors is the following: In calculating $\eta_F$ and $\eta_O$ in the leading order, we can safely replace the exact $\ket{\psi(t+\dt)}$ by a higher-order approximant $\ket{\psi_n(t+\dt)}$ for $n>m$.
Replacing $\ket{\psi(t+\dt)}$ by $\ket{\psi_n(t+\dt)}$ in $\eta_F$ and $\eta_O$, we obtain the following key analytical results (see Appendix~\ref{app:error} for derivation):
For the fidelity error,
\begin{align}
    \eta_F &= \eta_F^{(mn)} + O(\dt^{n+1}),\label{eq:etaF_mn}\\
    \eta_F^{(mn)} &\equiv \sqrt{1-|\braket{\psi_n(t+\delta t)|\psi_m(t+\delta t)}|^2},
\end{align}
and, for the observable error,
\begin{align}
    \eta_O &= \eta_O^{(mn)} + O(\dt^{n+1}),\label{eq:etaO_mn}\\
    \eta_O^{(mn)} &\equiv \braket{\psi_n(t+\delta t)|O|\psi_n(t+\delta t)} \notag\\ 
    &\qquad -\braket{\psi_m(t+\delta t)|O|\psi_m(t+\delta t)}.
\end{align}
Given that $\eta_F=O(\dt^{m+1})$ and $\eta_O=O(\dt^{m+1})$, these results mean that $\eta_F^{(mn)}$ ($\eta_O^{(mn)}$) coincides with $\eta_F$ ($\eta_O$) in the leading order since $m<n$.
If $n\le m$, the above equations hold true, but $O(\dt^{n+1})$ contributions are nonnegligible, and $\eta^{(mn)}$ do not give good estimates for $\eta$.}

Unlike $\eta_F$ and $\eta_O$, $\eta_F^{(mn)}$ and $\eta_O^{(mn)}$ consist of $T_m(\dt)$ and $T_n(\dt)$ and are thereby implementable in quantum circuits.
In other words, we can estimate the deviation from the exact solution induced by $T_m(\dt)$ without knowing the solution when supplemented with the fourth-order Trotterization and neglect higher-order corrections.
{\hl We will discuss in more detail how to measure $\eta_F^{(mn)}$ and $\eta_O^{(mn)}$ in Sec.~\ref{sec:eta_circuit}.}

We emphasize that $\eta_F$ and $\eta_O$ are the actual Trotter error specific to the current state $\ket{\psi(t)}$.
This contrasts the upper-bound arguments on the operator difference $U(\dt)-T_m(\dt)$~\cite{Kivlichan2020,Childs2021}.
Such upper bounds apply to arbitrary states and are thus always larger than or equal to the error occurring at a specific state $\ket{\psi(t)}$.
The fact that $\eta_F$ and $\eta_O$ are state-dependent enables us to choose $\dt$ more accurately so that the error is below our tolerance, as we will see in detail below.

{\hl 
We highlight two sets of $(m,n)$ of particular interest in practical use.
The first choice is the minimum pair $(m,n)=(1,2)$, for which the Trotterizations $T_m$ and $T_n$ involve the minimum possible exponentials, i.e., the gate complexity in the quantum circuit.
For noisy circuits, such lowest-order Trotterizations are commonly used to avoid gate errors as much as possible.
The second choice is the pair of minimum even numbers $(m,n)=(2,4)$, which can be useful when we can use more gates to achieve higher accuracy.
Changing $m$ from 1 to 2 increases the number of exponentials in $T_m$ from 2 to 3, by which the Trotter error $\eta_F$ or $\eta_O$ becomes one-order smaller.
In this case, using $n=4$ rather than $n=3$ could be beneficial.
To see this, let us compare Ruth's third-order formula~\cite{Ruth1983,Hatano2005},
\begin{align}
    T_3(\delta t) &\equiv e^{-i\frac{7}{24}A\delta t}e^{-i\frac{2}{3} B\delta t}e^{-i\frac{3}{4}A\delta t}e^{i\frac{2}{3}B\delta t}\notag\\
    &\qquad \times e^{i\frac{1}{24}A\delta t}e^{-i B\delta t}\label{eq:defT3}
\end{align}
and
the fourth-order Forest-Ruth-Suzuki (FRS) formula~\cite{Forest1990,Suzuki1990}
\begin{align}
    T_4(\delta t) &\equiv e^{-i\frac{s}{2}A\delta t}e^{-is B\delta t}e^{-i\frac{1-s}{2}A\delta t}e^{-i(1-2s)B\delta t}\notag\\
    &\qquad \times e^{-i\frac{1-s}{2}A\delta t}e^{-is B\delta t}e^{-i\frac{s}{2}A\delta t},\label{eq:defT4}
\end{align}
where $s=(2-2^{1/3})^{-1}$.
Here we note that the FRS formula involves only one extra exponential but is one-order more accurate than Ruth's third-order formula.
Since $\eta^{(24)}$ estimates $\eta^{(23)}$ one-order more accurately at the expense of an extra exponential, it can be worth using $n=4$ for $m=2$.
}

{\hl
\subsection{Quantum circuit implementations}\label{sec:eta_circuit}
In Sec.~\ref{sec:eta_eqs} we derived two expressions, Eqs.~\eqref{eq:eta_F_leading} and \eqref{eq:etaF_mn}, for estimating the fidelity error $\eta_F$ and one~\eqref{eq:etaO_mn} for the observable error $\eta_O$. We are assuming a quantum advantage regime, where the number of qubits is so large that $\ket{\psi(t)}$ cannot be stored in classical computer memory but is realized on a quantum circuit. In such a regime, all the expressions for $\eta_F$ and $\eta_O$ cannot be evaluated with classical linear algebraic computations, and we need to develop ways to evaluate them on quantum circuits.

First, let us consider how to measure the leading-order contribution in Eq.~\eqref{eq:eta_F_leading} of the fidelity error $\eta_F$, assuming that $\ket{\psi(t)}$ is realized on a quantum circuit accurately enough. For concreteness we focus on the first-order Trotterization $m=1$, for which $T_1(\dt)=e^{-iA\dt}e^{-iB\dt}$ and $i\Upsilon_2=-i[A,B]\dt^2$.
Since $i\Upsilon_2$ and $(i\Upsilon_2)^2$ are Hermitian, one can evaluate the expectation values for them based on samplings (we will discuss the sampling cost below). 

Although this method is useful for small quantum systems, it becomes quadratically costly for larger systems. To see this we consider a Hamiltonian $H$ with $A$ and $B$ are linear combinations of $L$ distinct local Pauli strings. Thanks to the commutator scaling~\cite{Childs2021}, $i\Upsilon_2$ consists of $O(L)$ distinct Pauli strings, and its expectation values are evaluated by estimating the expectation values of each Pauli string.
Notably, the translation symmetries can reduce the number of combinations of measurements for expectation value estimations, even down to $O(1)$.
However, $(i\Upsilon_2)^2$ is no more a commutator and involves $O(L^2)$ Pauli strings. In most cases symmetries cannot reduce the number of measured operators to be subextensive, and the sampling cost tends to be extensive.
This extensive sampling cost also appears in the energy variance estimation in Ref.~\cite{Zhao2022,Zhao2023}.
Thus Eq.~\eqref{eq:eta_F_leading} tends to be useful only in small quantum systems.

In contrast, $\eta_F^{(mn)}$ in Eq.~\eqref{eq:etaF_mn} avoids measuring numerous distinct Pauli strings.
To show this we write $\ket{\psi(t)}$ as $\ket{\psi(t)}=U_T(t)\ket{\psi(0)}=U_T(t)U_\mathrm{prep}\ket{\bm{0}}$.
Here $\ket{\bm{0}}$ denotes the initialized state, $U_\mathrm{prep}$ is an initial state preparation unitary, and $U_T(t)$ is some Trotterized unitary propagation from time 0 to $t$. We assume $U_\mathrm{prep}$ and $U_T(t)$ have appropriate circuit realizations.
With these notations $\eta_F^{(mn)}$ can be obtained by
\begin{align}
\eta_F^{(mn)}&=\sqrt{1-p_{\bm{0}} },\\
p_{\bm{0}}&\equiv
|\braket{\bm{0}|U_\mathrm{prep}^\dag U_T^\dag(t) T_n^\dag(\dt) T_m(\dt) U_T(t)|\bm{0}} |^2.
\end{align}
Our assumptions tell us that $U_\mathrm{prep}^\dag U_T^\dag(t) T_n^\dag(\dt) T_m(\dt) U_T(t)$ has a circuit realization, so $p_{\bm{0}}$ can be interpreted by the probability of finding $\ket{\bm{0}}$ after $\ket{\bm{0}}$ being evolved by the circuit (see also the middle panel of Fig.~\ref{fig:outline}).
Note that $p_{\bm{0}}$ is a kind of the Lochmidt echo~\cite{Wisniacki2012}.
A challenge is that $p_{\bm{0}}$ becomes exponentially small when the system size $L$ increases, so this procedure makes practical sense only in small systems.
This issue is not technical but intrinsic in the fidelity error because it is the most stringent quantifier among many-body wave function errors.

The observable error $\eta_O$ is more useful for extensive quantum systems. Typically we are interested in local Pauli strings or their linear combinations of $O(L)$. The expectation value of each Pauli string can be estimated by measuring the wave function $\ket{\psi(t)}$ in its eigenbasis. Unlike fidelity, the expectation value does not decay exponentially with the system size. Also, symmetries, such as the translation one, can reduce the number of measured strings to possibly $O(1)$. When one is only interested in local observables, using $\eta_O$ is the cheapest option to guarantee precision.

In each method, the Trotter error estimation is based on sampling, and the statistical error can be a bottleneck to achieve high accuracy. The statistical error in general scales as $\propto \mathcal{N}^{-1/2}$, with $\mathcal{N}$ denoting the number of samples (or shots). For the error estimation to be successful, this error should be smaller enough than the error, $\eta_F$ or $\eta_O$, so $\dt^{m+1}\gtrsim \mathcal{N}^{-1/2}$ must hold true. When we use $\eta^{(mn)}$ as an estimator, the estimation error is determined by the larger of $\sim\mathcal{N}^{-1/2}$ and $\sim \dt^{n+1}$ (see Eqs.~\eqref{eq:etaF_mn} and \eqref{eq:etaO_mn}). So, one cannot make the error estimation infinitely accurate by increasing $n$ when the available number of shots $\mathcal{N}$ is limited finitely. Rather, $\dt^{n+1}\sim \mathcal{N}^{-1/2}$ would hold for a reasonably chosen set of $\dt$, $n$, and $\mathcal{N}$. In Section~\ref{sec:finiteShots}, we will benchmark our way to estimate $\eta_O$ for $(m,n)=(2,4)$ in an example spin system and show that it works with $\mathcal{N}\sim10^5 (10^7)$ for $\epsilon_O=10^{-2} (10^{-3})$.
}

{\hl 
\section{Precision-guaranteed Trotterization}\label{sec:iteration}
In the previous section, we developed the methods to evaluate Trotter errors on quantum circuits.
Given that the Trotter error is known, one can make Trotterization precision-guaranteed: In each time step, one can make sure that the Trotter error is within a preset accuracy target $\epsilon$.
In this section we develop such an algorithm consisting of error measurements and step size optimization, as illustrated in Fig.~\ref{fig:outline} (The figure is for the fidelity version, but it works, in parallel, for the observable version.).

Our algorithm uses either $\eta_F^{(mn)}$ and $\eta_O^{(mn)}$ as the Trotter error estimator, and we name it Trotter$(m,n)$. For concreteness we set $(m,n)=(2,4)$ and describe Trotter24 since the generalization to other $(m,n)$ is straightforward (we may use Trotter$(m,n)$ and Trotter$mn$ interchangeably). In current NISQ devices, $(m,n)=(1,2)$ could be more realistic.
}
Since the argument goes in parallel, we first focus on the fidelity error and will address the observable error later in this section.

Our overall task is to simulate the time evolution according to the Hamiltonian $H$ from the initial time $t_\mathrm{ini}$ to the final time $t_\mathrm{fin}$, starting from an initial state $\ket{\psi_0}$.
We set an error tolerance $\epsilon$ for the fidelity error in each time step.
Initially, we have no a priori information about the appropriate time step, so take a reasonably small trial stepsize $\dt_0$, say, $\dt_0=0.1J^{-1}$ with $J$ being a typical energy scale of $H$.
One could also choose $\dt_0$ so small that $T_2(\dt_0)$ never gives larger error than our tolerance $\epsilon$, as guaranteed by a mathematical bound (see Eq.~\eqref{eq:dt_bound} and Appendix~\ref{sec:errorbound} for detail).
One can also use this $\dt_0$ to know the upper bound, in advance, for the required quantum resources for the calculation, although they tend to be too pessimistic, as we will see below.

For the trial $\dt_0$ taken in either way, we implement $T_2(\dt_0)$ and $T_4(\dt_0)$ and calculate $\eta_F^{(24)}$ using a quantum circuit.
Basically, we aim the stepsize to be so small that
\begin{align}\label{eq:etaF_ep}
\eta_F^{(24)}<\epsilon.
\end{align}
If this is true, we accept our trial $\dt_0$ and evolve our state as $\ket{\psi_2}=T_2(\dt_0)\ket{\psi_0}$.
If $\eta_F^{(24)}\ge\epsilon$ instead, our trial $\dt_0$ is too large and we need a smaller $\dt_0'$.
In choosing $\dt_0'$ appropriately, we invoke the leading-order scaling relation $\eta_F^{(24)}\approx \alpha\dt_0^3$ for some unknown $\alpha$ independent of $\dt_0$.
We can use this relation to estimate $\alpha$ by $\alpha\approx \eta_F^{(24)}/\dt_0^3$ since we measured $\eta_F^{(24)}$.
For $\dt_0'$, we expect $\eta_F^{(24)}{}'\approx \alpha (\dt_0')^3\approx \eta_F^{(24)}(\dt_0'/\dt_0)^3$, which we wish is smaller than $\epsilon$.
Thus, the condition $\eta_F^{(24)}{}'<\epsilon$ leads to $\dt_0'\approx \dt_0 (\epsilon /\eta_{F}^{(24)})^{1/3}$ as an optimal choice within our error tolerance.
For a safety margin, we introduce a constant $C$ ($0<C<1$) and set $\dt_0'= C \dt_0 (\epsilon /\eta_{F}^{(24)})^{1/3}$ as an updated trial $\dt_0$.
We repeat this update procedure until $\eta_F^{(24)}$ gets smaller than $\epsilon$ and accept the latest $\dt_0$ to evolve our state as $\ket{\psi_2}=T_2(\dt_0)\ket{\psi_0}$.  

Next, we move on to the second step, using a time step $\dt_1$.
In choosing this, we again use the latest $\eta_F^{(24)}$ obtained at the end of the previous time step.
Since $\ket{\psi_2}\approx \ket{\psi_0}$, we can expect the error scaling coefficient $\alpha$ to be almost the same in the present and previous steps.
Therefore, like in the updated trials within the previous time step, we have $\dt_1=C \dt_0 (\epsilon/\eta_F^{(24)})^{1/3}$ as a good candidate for the optimal stepsize in the present time step.
We note that $\eta_F^{(24)}$ here is what was measured in the previous step, and we have not made any measurements in the present step yet.
Using this $\dt_1$ as a trial stepsize, we implement $T_2(\dt_1)$ and $T_4(\dt_1)$ and calculate $\eta_F^{(24)}$ using a quantum circuit.
Depending on whether $\eta_F^{(24)}$ is less or greater than $\epsilon$, we accept or update $\dt_1$ like in the previous step.

The following iteration is straightforward and repeated until the accumulated evolution time $t_\mathrm{ini}+\dt_0+\dt_1+\dots$ exceeds the final time $t_\mathrm{fin}$.
We summarize a pseudocode for the algorithm in Algorithm~\ref{alg:fidelity}.

\begin{algorithm}
\caption{Fidelity-based Trotter24}\label{alg:fidelity}
\flushleft{
\textbf{Input:} Initial and final times, $t_\mathrm{ini}$ and $t_\mathrm{fin}$, an initial state $\ket{\psi_0}$, a Hamiltonian $H=A+B$, an error tolerance $\epsilon$, an initial stepsize $\delta t_0$, a safety constant $C$ ($0<C<1$), an oracle function $\mathrm{FIDELITY}(\ket{\phi},\ket{\psi})$ that calculates $|\braket{\phi|\psi}|^2$.\\[1mm]
\textbf{Output:} An ordered list of unitaries $U_\mathrm{list}$ that approximates $e^{-i H (t_\mathrm{fin}-t_\mathrm{ini})}$ within the error tolerance for each time step.
}
\begin{algorithmic}[1]
\State $t \gets t_\mathrm{ini}$
\State $\delta t \gets \delta t_0$
\State $U_\mathrm{list}=\{\}$ (empty list)
\While{$t+\delta t < t_\mathrm{fin}$}
    \State $\ket{\psi(t)}\gets \prod_{k} (U_\mathrm{list})_k\ket{\psi_0}$
    \Do
        \State $T_2(\delta t)\gets e^{-i A\delta t/2}e^{-i B\delta t}e^{-i A\delta t/2}$
        \State $T_4(\delta t)\gets e^{-i\frac{s}{2}A\delta t}$ $e^{-is B\delta t}$ $e^{-i\frac{1-s}{2}A\delta t}$ $e^{-i(1-2s)B\delta t}$ $\times e^{-i\frac{1-s}{2}A\delta t}$ $e^{-is B\delta t}$ $e^{-i\frac{s}{2}A\delta t}$
        \State $\eta \gets 1-\mathrm{FIDELITY}(T_4(\delta t)\ket{\psi(t)},T_2(\delta t)\ket{\psi(t)})$
        \State $\delta t\gets C\cdot (\epsilon/\eta)^{1/3}\delta t$
    \doWhile{$\eta > \epsilon$}
    \State Prepend $T_2(\delta t)$ to the ordered list $U_\mathrm{list}$
    \State $t\gets t+\delta t$
\EndWhile
\Return $U_\mathrm{list}$
\end{algorithmic}
\end{algorithm}

Let us make a parallel argument for the observable error $\eta_O$ instead of the fidelity error $\eta_F$.
At each time step, we measure $\eta_O^{(24)}$ and judge if the condition
\begin{align}
    |\eta_O^{(24)}| < \epsilon_O \| O\|
\end{align}
is met.
This is an analog of Eq.~\eqref{eq:etaF_ep}, and we introduced the operator norm $\|O\|$ as a reference scale and put the subscript $O$ on the tolerance as $\epsilon_O$ to avoid confusion.
The iteration scheme is parallel to the fidelity case.
We summarize a pseudocode for the observable-based algorithm in Algorithm~\ref{alg:observable}.

\begin{algorithm}
\caption{Observable-based Trotter24}\label{alg:observable}
\flushleft{
\textbf{Input:} Initial and final times, $t_\mathrm{ini}$ and $t_\mathrm{fin}$, an initial state $\ket{\psi_0}$, a Hamiltonian $H=A+B$, an error tolerance $\epsilon_O$, an initial stepsize $\delta t_0$, a safety constant $C$ ($0<C<1$), an oracle function $\mathrm{EXP}(O,\ket{\psi})$ that calculates $\braket{\psi|O|\psi}$.\\[1mm]
\textbf{Output:} An ordered list of unitaries $U_\mathrm{list}$ that approximates $e^{-i H (t_\mathrm{fin}-t_\mathrm{ini})}$ within the error tolerance for each time step.
}
\begin{algorithmic}[1]
\State $t \gets t_\mathrm{ini}$
\State $\delta t \gets \delta t_0$
\State $U_\mathrm{list}=\{\}$ (empty list)
\While{$t+\delta t < t_\mathrm{fin}$}
    \State $\ket{\psi(t)}\gets \prod_{k} (U_\mathrm{list})_k\ket{\psi_0}$
    \Do
        \State $T_2(\delta t)\gets e^{-i A\delta t/2}e^{-i B\delta t}e^{-i A\delta t/2}$
        \State $\ket{\psi_2}\gets T_2(\delta t)\ket{\psi(t)}$
        \State $T_4(\delta t)\gets e^{-i\frac{s}{2}A\delta t}$ $e^{-is B\delta t}$ $e^{-i\frac{1-s}{2}A\delta t}$ $e^{-i(1-2s)B\delta t}$ $\times e^{-i\frac{1-s}{2}A\delta t}$ $e^{-is B\delta t}$ $e^{-i\frac{s}{2}A\delta t}$
        \State $\ket{\psi_4}\gets T_4(\delta t)\ket{\psi(t)}$
        \State $\eta \gets |\mathrm{EXP}(O,\ket{\psi_4})-\mathrm{EXP}(O,\ket{\psi_2})|$
        \State $\delta t\gets C\cdot (\epsilon/\eta)^{1/3}\delta t$
    \doWhile{$|\eta| > \epsilon_O \|O\|$}
    \State Prepend $T_2(\delta t)$ to the ordered list $U_\mathrm{list}$
    \State $t\gets t+\delta t$
\EndWhile
\Return $U_\mathrm{list}$
\end{algorithmic}
\end{algorithm}

\vspace{3mm}
\section{Error propagation}
{\hl
In this section we analyze the fidelity and observable Trotter errors in multiple Trotter steps, showing that both errors increase at most linearly in the number of steps. For generality we consider an $m$-th order Trotterization, which was set to be $m=2$ in the previous section.
}

After $N$ $(\ge1)$ steps, we obtain a quantum state
\begin{align}
    \ket{\psi_m(t_N)} = \prod_{i=0,\dots,N-1}^{\leftarrow} T_m(\dt_i)\ket{\psi_0}
\end{align}
at time
\begin{align}
    t_N = t_\mathrm{ini}+\sum_{i=0}^{N-1}\dt_i
\end{align}
as an approximation for the exact state
\begin{align}
    \ket{\psi(t_N)} = \prod_{i=0,\dots,N-1}^{\leftarrow} U(\dt_i)\ket{\psi_0} = e^{-i H (t_N -t_\mathrm{ini})}\ket{\psi_0}.
\end{align}
This section gives upper bounds for accumulated errors in the $N$ steps.
As expected, we will have error propagation linear in $N$.
Throughout this section, we let $\sim$ and $\lesssim$ denote $=$ and $\le$, respectively, when sub-leading terms in $\dt_j$ $(j=0,\dots,N-1)$ are neglected.

For the fidelity-based Trotter$(m,n)$, 
let us find an upper bound for the accumulated error
\begin{align}
    \eta_{F,N} &\equiv \sqrt{1- \left| \braket{\psi(t_N)|\psi_m(t_N)}\right|^2}.\label{eq:upper1}
\end{align}
As we derive in Appendix~\ref{app:error_propagation}, 
\begin{align}\label{eq:Fupper}
    \eta_{F,N} \lesssim  N \epsilon,
\end{align}
meaning a linear increase in the fidelity error.
This upper bound implies an upper bound for the error in the expectation value of an arbitrary observable $O$,
\begin{align}\label{eq:errorO}
    \eta_{O,N} \equiv \left| \braket{\psi(t_N)|O|\psi(t_N)} - \braket{\psi_m(t_N)|O|\psi_m(t_N)} \right|.
\end{align}
To derive the bound for $\eta_{O,N}$, we recall $\eta_{O,N} \le 2 D(\ket{\psi(t_N)}\bra{\psi(t_N)},\ket{\psi_m(t_N)}\bra{\psi_m(t_N)}) \|O\|$, where $D(\rho,\sigma)$ denotes the trace distance, which is known to satisfy $D(\rho,\sigma)\le \sqrt{1-F(\rho,\sigma)}$ with $F(\rho,\sigma)$ being the fidelity.
Using Eq.~\eqref{eq:Fupper}, we obtain
\begin{align}\label{eq:boundO1}
    \eta_{O,N} \lesssim N\epsilon\|O\|,
\end{align}
which linearly increases with $N$.

For the observable-based Trotter$(m,n)$ using an observable $O$, let us find an upper bound with the tolerance $\epsilon_O$ for the accumulated error~\eqref{eq:errorO} arising in the same observable.
We can prove
\begin{align}\label{eq:boundO2}
    \eta_{O,N} \lesssim N \epsilon_O \|O\|
\end{align}
by induction on $N$.
This claim trivially holds when $N=1$ by the definition of $\dt_0$.
Suppose that Eq.~\eqref{eq:boundO2} for $N$.
For $N+1$, the triangle inequality gives
\begin{widetext}
\begin{align}
    \eta_{O,N+1} &\le
    \left| \braket{\psi(t_{N+1})|O|\psi(t_{N+1})} - \braket{\psi_m(t_N)|U^\dag(\dt_{N})OU(\dt_{N})|\psi(t_N)}  \right|\notag\\
    &\qquad+\left| \braket{\psi_m(t_N)|U^\dag(\dt_{N})OU(\dt_{N})|\psi(t_N)} - \braket{\psi_m(t_{N+1})|O|\psi_m(t_{N+1})} \right|.
\end{align}
\end{widetext}
Since $U^\dag(\dt_{N})OU(\dt_{N})\sim O$, 
the first term on the right-hand side is $\lesssim N\epsilon_O\|O\|$ by the inductive hypothesis.
The second term on the right-hand side is less than $\epsilon_O \|O\|$, as implemented in the algorithm.
Combining these two, we obtain $\eta_{O,N+1}\lesssim (N+1)\epsilon_O\|O\|$, which completes the induction.

\section{benchmark implementations}\label{sec:benchmark}
{\hl 
In this section we implement and benchmark Trotter$(m,n)$, using a classical computer. In the first subsection we consider the ideal limit of the infinite number of shots, for which the error estimators $\eta_F^{(mn)}$ and $\eta_O^{(mn)}$ can be obtained without statistical errors. Then, in the second subsection, we discuss a more realistic situation, where the number of shots is limited to be finite. In both subsections we neglect errors in the quantum circuit.

\subsection{Ideal limit of $\mathcal{N}_\mathrm{shots}=\infty$}\label{sec:infShots}
}

\begin{figure}
    \centering
    \includegraphics[width=\columnwidth]{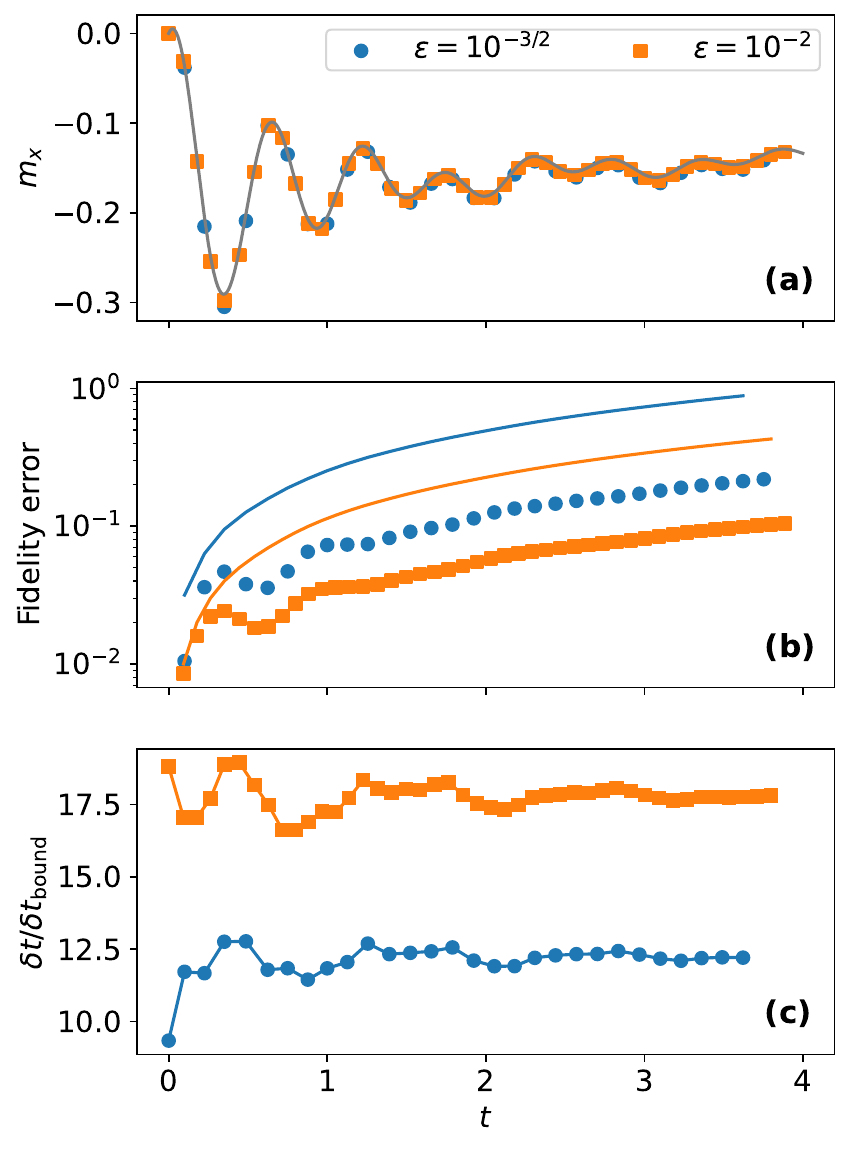}
    \caption{(a) Dynamics of $x$-magnetization density calculated by fidelity-based Trotter24 for tolerance $\epsilon=10^{-3/2}$ (circle) and $10^{-2}$ (square). The solid curve shows the accurate solution, the system size is $L=18$, and the safety constant is $C=0.95$.
    (b) The actual fidelity errors $\eta_F$~\eqref{eq:def_error_F} in the simulation presented in panel (a). The solid curves show their upper bounds~\eqref{eq:Fupper}. Blue (orange) points and curve correspond to the case of $\epsilon=10^{-3/2}$ ($10^{-2}$).
    (c) The ratio of the stepsize $\dt$ chosen in each step to $\dt_\mathrm{bound}$ obtained by the error-bound approach~\eqref{eq:dt_bound}. Different symbols correspond to those in panel (a).}
    \label{fig:demo}
\end{figure}

First, we implement the fidelity-based Trotter24 and will address the observable-based one later in this section. {\hl In both cases we assume that both $\eta_F^{(mn)}$ and $\eta_O^{(mn)}$ can be obtained without statistical errors.}
Following Ref.~\cite{Zhao2022}, we consider the following Hamiltonian
\begin{align}
\begin{split}
    A = h_x \sum_{j=1}^L \sigma^x_j,\quad 
    B = \sum_{j=1}^L (J_z \sigma^z_j \sigma^z_{j+1} + h_z \sigma^z_j),
\end{split}\label{eq:Hpm}
\end{align}
where $\sigma_j^\alpha$ are the Pauli matrices acting on the $j$-th site, periodic boundary conditions are imposed, and we set $J_z=-1.0$, $h_z=0.2$, and $h_x=-2.0$.
Taking the initial state fully polarized along the $-y$ direction, we let it evolve for a while.
Figure~\ref{fig:demo}(a) shows the expectation value of the $x$-magnetization density
\begin{align}
    m_x \equiv \frac{1}{L}\sum_{j=1}^L \sigma_j^x
\end{align}
for different tolerances $\epsilon=10^{-3/2}$ and $10^{-2}$ at $L=18$, and we set $C=0.95$.
As expected, for smaller tolerance, the simulated dynamics resemble the exact result better, as shown in the upper panel.
We note that we encounter few $\eta^{(24)}>\epsilon$ in these simulations; It happens three (five) times for $\epsilon=10^{-3/2}$ ($10^{-2}$) during the simulation time range.
These numbers further decrease as we decrease the safety constant $C$, as we will discuss below.
Figure~\ref{fig:demo}(b) shows the actual fidelity error $\eta_F$ in those simulations.
We confirm that the errors are well below the upper bound~\eqref{eq:Fupper}, especially in late times.

The adaptively chosen stepsize $\dt$ is significantly larger than the one obtained by the error-bound approach.
According to Ref.~\cite{Kivlichan2020}, their tight error bound of the second-order product formula gives the possible maximum stepsize (see Appendix~\ref{sec:errorbound} for more detail)
\begin{align}\label{eq:dt_bound}
    \dt_\mathrm{bound} = \left(\frac{\epsilon}{\| [B,[B,A]]\| + \frac{1}{2}\| [A,[B,A]]\|}\right)^{1/3},
\end{align}
for which the difference between $U(\dt)$ and the second-order Trotterization does not exceed the tolerance $\epsilon$.
As shown in Fig.~\ref{fig:demo}(c), the ratio of the adaptively chosen $\dt$ to $\dt_\mathrm{bound}$ is roughly greater than 10.
This means that the stepsize determined by the error bound tends to be too small for a given tolerance, and the adaptive stepsize is significantly larger.
This discrepancy derives from the fact that Trotter24 utilizes the quantum state at each time step while the error bound applies to arbitrary states and tends to be too pessimistic.

\begin{figure}
    \centering
    \includegraphics[width=\columnwidth]{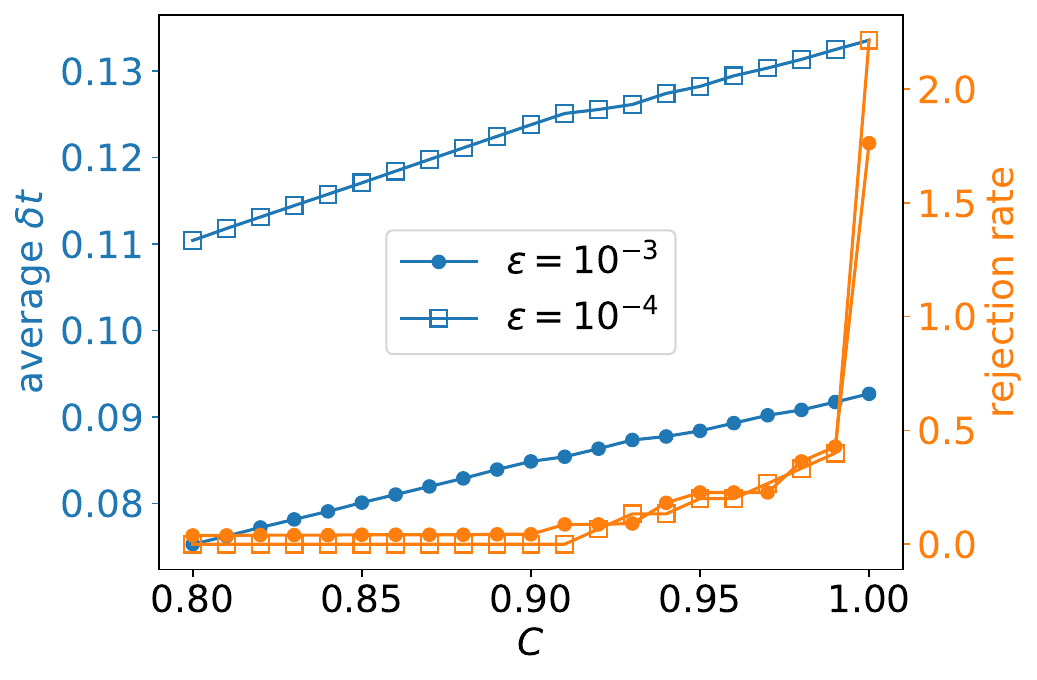}
    \caption{The $C$-dependence of the average stepsize $\dt$ (left $y$-axis) and the rejection rate (right $y$-axis), i.e., the occurrence of $\eta^{(24)}_F>\epsilon$ per time step.
    Different symbols correspond to $\epsilon=10^{-3/2}$ (circle) and $10^{-2}$ (square).
    The averages are taken for $t_\mathrm{ini}=0.0$ and $t_\mathrm{fin}=2.0$ with the initial trial stepsize $\dt_0=0.1$.}
    \label{fig:Cdep}
\end{figure}

The $C$-dependence of the algorithm is shown in Fig.~\ref{fig:Cdep}.
For various $C$ ($0.8\le C\le 1.0$), we run the Trotter24 with the other parameters being the same as in Fig.~\ref{fig:demo}.
Over the time interval $t_\mathrm{ini}=0.0$ and $t_\mathrm{fin}=4.0$, we measure the average of the adopted stepsize $\dt$ and the rejection rate, i.e., the average number of occurrences of $\eta^{(24)}_F>\epsilon$ per each time step.
As the left $y$-axis shows, the average stepsize is nearly proportional to $C$, as expected from its definition.
Meanwhile, the rejection rate increases only slowly as $C$ increases, except for the close vicinity of $C=1$ ($0.99\lesssim C\le 1$), where $C$ rapidly increases to exceed unity.
If we never mind repeatedly measuring $\eta^{(24)}_F$, the choice $C=1$ is ideal for making $\dt$ larger, i.e., the circuit depth shallower.
However, by choosing a slightly smaller $C$, like $C=0.95$ or $0.90$, we benefit from a dramatically-reduced rejection rate in exchange for a slight increase in the circuit depth.

\begin{figure}
    \centering
    \includegraphics[width=\columnwidth]{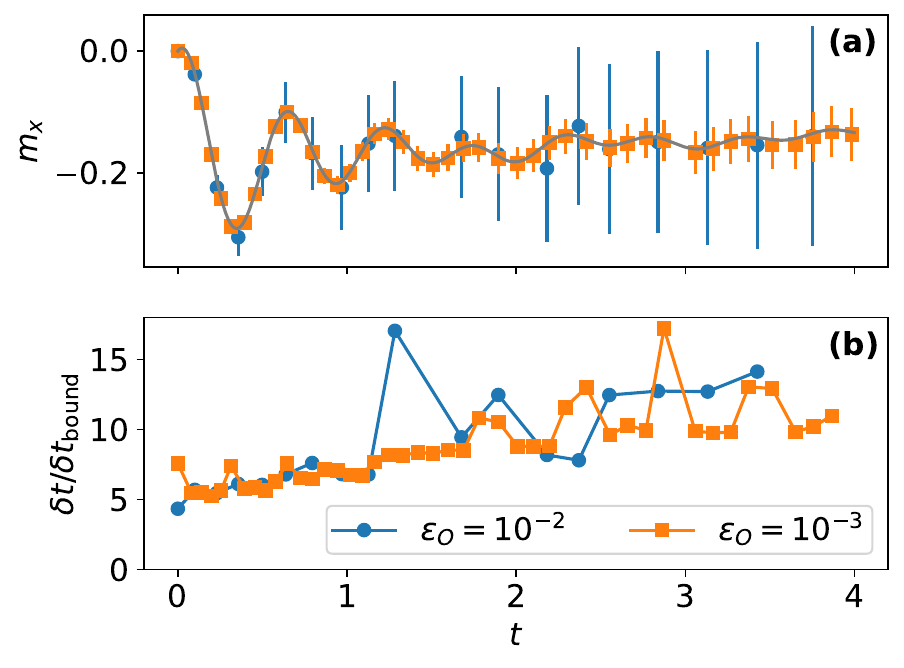}
    \caption{(a) Dynamics of $x$-magnetization density $m_x$ under the Hamiltonian~\eqref{eq:Hpm} calculated by observable-based Trotter24 for $O=m_x$ and tolerance $\epsilon_O=10^{-2}$ (circle) and $10^{-3}$ (square). The error bar shows the range where the exact solution resides as indicated by the theoretical upper bound~\eqref{eq:boundO2}. The solid curve shows an accurate solution obtained by a small enough $\dt$, the system size is $L=18$, and the safety constant is $C=0.95$.
    (b) The ratio of the stepsize $\dt$ chosen in each step to $\dt_\mathrm{bound}$ obtained by the error-bound approach~\eqref{eq:dt_bound} in the simulations shown in panel (a).}
    \label{fig:demo_obs}
\end{figure}

Now we implement the observable-based Trotter24 for the same model~\eqref{eq:Hpm} in the same setup.
Suppose again that we are interested in simulating the dynamics of $m_x$.
For this purpose, it is natural to set $O=m_x$, for which the Trotter24 generates Fig.~\ref{fig:demo_obs}.
For the smaller tolerance $\epsilon_O=10^{-3}$, we obtain more accurate results for $\braket{ \psi(t)|m_x |\psi(t)}$, as expected.
As shown in panel (b), the stepsize is at least 5 times larger than $\dt_\mathrm{bound}$ given by Eq.~\eqref{eq:dt_bound}, whose values are $\dt_\mathrm{bound}=2.31\times10^{-2}$ for $\epsilon=10^{-2}$ and $1.07\times10^{-2}$ for $\epsilon=10^{-3}$.
We note that the exact value always resides within the error bar representing the theoretical upper bound~\eqref{eq:boundO2}.
Even when the exact values are not available, the upper bound tells us in what region they are.

Before closing this section, we remark on the stability and efficiency of Trotter24 for simulations over reasonably long times, as seen in Fig.~\ref{fig:demo}(a).
This is an advantage over extrapolation methods, such as Richardson's~\cite{Endo2018}, in which physical quantities like $\braket{\psi(t)|m_x|\psi(t)}$ are obtained by extrapolating their estimates using different Trotter steps.
The extrapolation methods are particularly useful for short times $t\ll 1$ (in units of an inverse local energy scale) because $\braket{\psi(t)|m_x|\psi(t)}$ is well approximated by a low-order polynomial in the Trotter steps and allows us to extrapolate the exact solution as the infinite steps limit~\cite{Endo2019}.
For $t\gg1$, however, the required steps increase significantly, and the estimates with limited steps tend to be unstable due to Runge's phenomenon (see Appendix~\ref{sec:extrapolation} for demonstration).
Although this pathologic behavior has been addressed using quantum singular value transformations~\cite{rendon2022improved}, it requires a fault-tolerant quantum computer.
When compared under the same gate complexity, Trotter24 is more stable than extrapolation methods, as detailed in Appendix~\ref{sec:extrapolation}.

{\hl
\subsection{Effects of finite $\mathcal{N}_\mathrm{shots}$}\label{sec:finiteShots}
In Section~\ref{sec:infShots}, we assumed that expectation values are exactly obtained without statistical errors, demonstrating the idealistic behavior of Trotter$(m,n)$. There the only source of error was the Trotter error. In reality, however, the number of available shots (i.e., the number of measurements of circuits) is limited finitely. Thus, in order to guarantee the accuracy of simulations, we need to make sure both the statistical error in the observable evaluation and the Trotter error are within our tolerance. In this subsection we discuss how to implement Trotter$(m,n)$ to control the Trotter error. 
We consider the same setup as in Sec.~\ref{sec:infShots} and aim to estimate $\mathcal{N}_\mathrm{shots}$ necessary in obtaining the time evolution of $O=m_x$ in the time interval $[t_\mathrm{ini},t_\mathrm{fin}]$ within tolerance $\epsilon_O$.

Before discussing Trotter24, we consider the conventional constant-step (second-order) Trotterization approach. For a fixed stepsize $\dt$, we compute $\ket{\psi_2(N\dt)}=T_2(\dt)^N \ket{\psi_0}$ for an $N$-step evolution. For each step $N$, we are computing the desired expectation value
\begin{align}
    &\braket{\psi_2(N\dt) | O | \psi_2(N\dt)}\notag\\
    &= \frac{1}{L}\sum_{j=1}^L  \braket{\psi_2(N\dt) | \sigma_j^x | \psi_2(N\dt)}\\
    &= \frac{1}{L}\sum_{j=1}^L  \braket{\psi_2(N\dt) |H_j \sigma_j^z H_j| \psi_2(N\dt)}\\
    &= \sum_{\bm{z}} \frac{1}{L}\sum_{j=1}^L z_j P(\bm{z})
    =\frac{1}{L}\sum_{j=1}^L \sum_{z_j = \pm 1} z_j P_j(z_j),  \label{eq:Pz}
\end{align}
where $H_i$ is the Hadamard gate acting on site $j$, $P(\bm{z})=|\braket{ \bm{z} | H_j| \psi_2(N\dt)}|^2$ denotes the probability of finding $\bm{z}=(z_1,z_2,\dots,z_L)$ with $z_j = 1-2b_j$ and $b_j=0,1$, and $P_j(z_j) = \sum_{\bm{z}\backslash z_j}P(\bm{z})$ is the marginal probability for site $j$. Equation~\eqref{eq:Pz} allows us to evaluate the expectation value based on sampling, and the statistical error of estimating the set of marginal probabilities $P_1,\dots,P_L$ is $\sim \mathcal{N}_\mathrm{meas}^{-1/2}$ with $\mathcal{N}_\mathrm{meas}$ being the number of measurements for each time step $N$. If we take account of the statistical independence among $P_j$'s, the statistical error of $\langle O\rangle$ is reduced to be $\sim(L\mathcal{N}_\mathrm{meas})^{-1/2}$. Thus, to make sure the estimation error is within the target accuracy with the $p$-$\sigma$ confidence interval $(p>0)$, we impose $p(L \mathcal{N}_\mathrm{meas})^{-1/2} < \epsilon_O$, which means that the minimum required number of measurements is 
\begin{align}\label{eq:Nmeas0}
    \mathcal{N}_\mathrm{meas}^0=\left\lceil \left(\frac{p}{\epsilon_O}\right)^2/L \right\rceil,
\end{align}
where $\lceil \cdots \rceil$ denotes the ceiling function.
For instance, if we demand 2-$\sigma$ confidence interval for $\epsilon_O=1\%$, $\mathcal{N}_\mathrm{meas}^0 = 4\times 10^4/L$. As shown below, this number is smaller than those for Trotter24 and the ADA Trotter~\cite{Zhao2022}. On the other hand, a challenge in the conventional Trotterization is choosing appropriate stepsize $\dt$ so that the Trotter error is within our tolerance $\epsilon_O$. As discussed in Sec.~\ref{sec:benchmark}, $\dt$, if chosen based on rigorous bounds, tends to be too small, and the number of steps $N_\mathrm{step}\approx (t_\mathrm{fin}-t_\mathrm{ini})/\dt$ and hence the circuit depth become too large. If one measures $O=m_x$ at every step, the total number of shots becomes $\mathcal{N}_\mathrm{shots}=N_\mathrm{step}\mathcal{N}_\mathrm{meas}^0$.

The number of measurements at each step that Trotter24 requires is twice as large as that the conventional Trotterization does because the Trotter error estimator $\eta_O^{(24)}$ involves two expectation values (see Eq.~\eqref{eq:etaO_mn}). Thus  $\mathcal{N}_\mathrm{shots}=2N_\mathrm{step}'\mathcal{N}_\mathrm{meas}^0$, where $N_\mathrm{step}'$ is the number of steps for Trotter24 and tends to be smaller than $N_\mathrm{step}$. The factor 2 appears This statistical error contributes to $\eta_O^{(24)}$ by $(2L\mathcal{N}_\mathrm{meas}^0)^{-1/2}\|O\|$, where the factor 2 again takes account of the two expectation values for the second- and fourth-order Trotterizations in $\eta_O^{(24)}$.
and the criterion $|\eta_O^{(24)}| < \epsilon_O \| O\|$ should be modified as $|\eta_O^{(24)}| +  (2L\mathcal{N}_\mathrm{meas}^0)^{-1/2}\|O\| < \epsilon_O \| O\|$ to guarantee precision within the $p$-$\sigma$ confidence interval. Equation~\eqref{eq:Nmeas0} simplifies this inequality as
\begin{align}\label{eq:error_p}
    |\eta_O^{(24)}| < \left(1-\frac{1}{\sqrt{2}p}\right)\epsilon_O\|O\|.
\end{align}
We note that the formulation reduces to the ideal case in Sec.~\ref{sec:infShots} in the limit of $p\to\infty$, where $\mathcal{N}_\mathrm{meas}^0\to \infty$ and Eq.~\eqref{eq:error_p} becomes $|\eta_O^{(24)}|<\epsilon_O \| O\|$.

\begin{figure}
    \centering
    \includegraphics[width=\columnwidth]{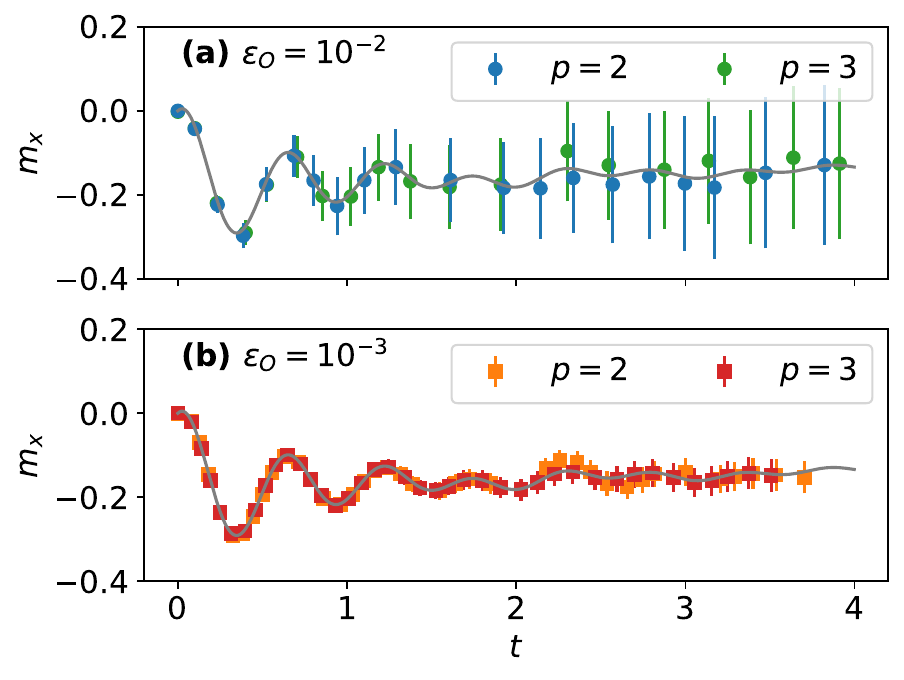}
     \includegraphics[width=\columnwidth]{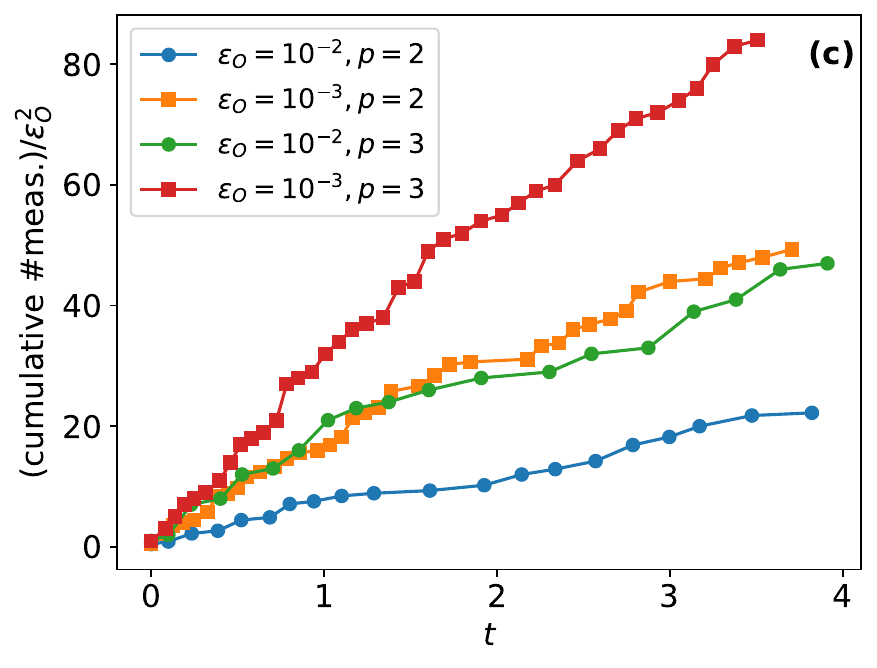}
    \caption{{\hl (a,b) Dynamics of $x$-magnetization density $m_x$ under the Hamiltonian~\eqref{eq:Hpm} calculated by observable-based Trotter24 for $O=m_x$ and tolerance (a) $\epsilon_O=10^{-2}$ and (b) $10^{-3}$, where the expectation values are calculated by sampling with $p=2$ (circle) and $p=3$ (square) (see text for detail). The error bar shows the range where the exact solution resides as indicated by the theoretical upper bound~\eqref{eq:boundO2}, which is here interpreted as the $p$-$\sigma$ confidence interval. The solid curve shows an accurate solution obtained by a small enough $\dt$, the system size is $L=18$, and the safety constant is $C=0.95$.
    (c) Cumulative number of measurements performed in obtaining panels (a) and (b) divided by $\epsilon_O^2$. }
    }
    \label{fig:demo_obs_shots}
\end{figure}

Figure~\ref{fig:demo_obs_shots} demonstrates how Fig.~\ref{fig:demo_obs} changes when we evaluate the expectation values based on sampling with the $p$-$\sigma$ confidence interval discussed above. The implementation is the same as in there except for the following two parts. First, the expectation values for the observable and $\eta_O^{(24)}$ are obtained by sampling using Eq.~\eqref{eq:Pz} and its counterpart for $\ket{\psi_4(N\dt)}$ with Eq.~\eqref{eq:Nmeas0}. Second, the error threshold is replaced by Eq.~\eqref{eq:error_p} in the $p$-dependent way. Note that the error bound~\eqref{eq:boundO2} is interpreted as the $p$-$\sigma$ confidence interval since the error estimation involves statistical errors. The Trotter error tends to decrease as $\epsilon_O$ decreases or $p$ increases, although such $p$ dependence is unclear for $t\gtrsim2$ in $\epsilon_O=10^{-2}$, where the error bar becomes very large.

The cumulative number of measurements before time $t$ is plotted in Fig.~\ref{fig:demo_obs_shots}(c). Considering Eq.~\eqref{eq:Nmeas0} representing the number of measurements at each evaluation of $\eta_O^{(24)}$ and $\langle O\rangle_t$, we have rescaled the number by $\epsilon_O^{-2}$ in the plot. Nicely, the number increases in time almost linearly since the rejection rate in finding an optimal $\dt$ in each step is kept low. Typical cumulative numbers of measurements for a unit of time $t=1$ are $10^{5}$ for $\epsilon_O=10^{-2}$ and $10^{7}$ for $\epsilon_O=10^{-3}$, which could be achievable in current NISQ devices.

Finally, we discuss the number of measurements required in the ADA Trotter~\cite{Zhao2022}, which adaptively chooses $\dt$ so that the energy expectation value and variance are close to their ideal values within tolerance. Unlike Trotter24 evaluating $O$, the ADA needs the expectation values of $H=A+B$ and $H^2 = (A+B)^2$. While $\langle H\rangle_t$ only requires a similar number of measurements to $\langle O\rangle_t$, computing $\langle H^2\rangle_t$ based on sampling is significantly more costly because $H^2$ is not necessarily local. For example, $H^2$ involves $AB+BA$, and it contains terms like $\sigma_{j'}^x \sigma_j^z\sigma_{j+1}^z$ ($1\le j,j'\le L$). Such terms cannot be simultaneously measured in a single circuit, and one needs multiple, at least $O(L)$, circuits for measurements. Thus, the number of shots required in the ADA Trotter is $O(L)$ greater than that in Trotter24 and the conventional Trotterization. Also, we emphasize again that the tolerance $\epsilon$ set for the energy expectation value and variance in the ADA Trotter cannot be translated to the error in the observable $O$ of interest for nonequilibrium states, and the precision guarantee is still elusive, unlike Trotter$(m,n)$.

}

\section{Generalization to Time-dependent Hamiltonians}\label{sec:tdep}
Trotter24, developed thus far for time-independent Hamiltonians, is straightforwardly generalized for time-dependent Hamiltonians unlike the previous study~\cite{Zhao2022}.
Their study is based on the energy conservation law, which is absent in time-dependent Hamiltonians, and its generalization to those Hamiltonians is not straightforward and has not been established yet.

Our setup is a generalization of Eq.~\eqref{eq:AandB} as
\begin{align}
    H(t) = A(t)+B(t),
\end{align}
and {\hl we consider approximating the exact evolution $\ket{\psi(t+\dt)}=U(t,\dt)\ket{\psi(t)}$ for $U(t,\dt) = \mathcal{T}\exp\left( -i \int_{t}^{t+\dt}H(s)ds\right)$, where $\mathcal{T}$ denotes the time ordering.}
Assuming that the quantum state $\ket{\psi(t)}$ at time $t$ is known, we try to approximate the subsequent time evolution for a stepsize $\dt$ by the socalled midpoint rule
\begin{align}
    \ket{\psi_2(t+\dt)} &= T_2(t,\dt)\ket{\psi(t)},\\
    T_2(t,\dt) &\equiv e^{-i A(t+\dt/2)\frac{\dt}{2}}e^{-i B(t+\dt/2)\dt}\notag\\
    &\qquad\qquad \times e^{-i A(t+\dt/2)\frac{\dt}{2}}.
\end{align}
The midpoint rule is known to be a second-order formula, and the fidelity error $\eta_F$ and the observable error $\eta_O$ are as small as $O(\dt^3)$, like in the time-independent-Hamiltonian cases.

To measure the errors in the leading order without using the exact state $\ket{\psi(t+\dt)}$, we use a fourth-order Trotterization formula for time-dependent Hamiltonians.
Focusing on a special case where $A(t)=a(t)A$ and $B(t)=b(t)B$ with $a(t)$ and $b(t)$ are scalars, we utilize the minimum fourth-order Trotterization formula~\cite{Ikeda2022}
\begin{align}
    \ket{\psi_4(t+\dt)} &= T_4(t,\dt)\ket{\psi(t)},\\
    T_4(t,\dt) &\equiv e^{(\frac{s \beta_1}{2}-u) A}e^{s \beta_2 B}e^{\frac{1-s}{2}\beta_1 A}e^{(1-2s)\beta_2 B} \notag\\
&\qquad\qquad \times e^{\frac{1-s}{2}\beta_1 A}e^{s \beta_2 B}e^{(\frac{s \beta_1}{2}+u) A}
\end{align}
consisting of seven exponentials, 
where $\beta_1 = \int_t^{t+\dt}a(s)ds$, $\beta_2=\int_t^{t+\dt}b(s)ds$, and $\beta_{12}=\frac{1}{2}\int_{t}^{t+\dt}dt_2 \int_{t}^{t_2}dt_1[b(t_2)a(t_1)-a(t_2)b(t_1)]$, and $u\equiv \beta_{12}/\beta_2$ is assumed to be $O(\dt^2)$.
For more general $A(t)$ and $B(t)$, one can utilize the fourth-order Suzuki formula~\cite{Hatano2005} consisting of 15 exponentials.
We can define $\eta^{(24)}_F$ and $\eta^{(24)}_O$ similarly to the case of time-independent Hamiltonians and confirm that $\eta_F\approx \eta_F^{(24)}$ and $\eta_O\approx \eta_O^{(24)}$ in their leading orders.

The algorithms of Trotter24 for time-dependent Hamiltonians are obtained by the replacements $T_2(\dt)\to T_2(t,\dt)$ and $T_4(\dt)\to T_4(t,\dt)$ in Algorithms~\ref{alg:fidelity} and \ref{alg:observable}.
{\hl Even though the time-ordered exponential $\mathcal{T}\exp$ complicates the propagator $U(t,\dt)$ for time-dependent Hamiltonians, Trotterizations $T_2(t,\dt)$ and $T_4(t,\dt)$ tailored for these cases correctly approximate it as $U(t,\dt)=T_2(t,\dt)+O(\dt^3)$ and $U(t,\dt)=T_4(t,\dt)+O(\dt^5)$.
The algorithms of Trotter24 are thus straightforwardly applied to time-dependent Hamiltonians because
the complication due to the time dependence is appropriately taken care of by Trotterization formulas.}

\begin{figure}
    \centering
    \includegraphics[width=\columnwidth]{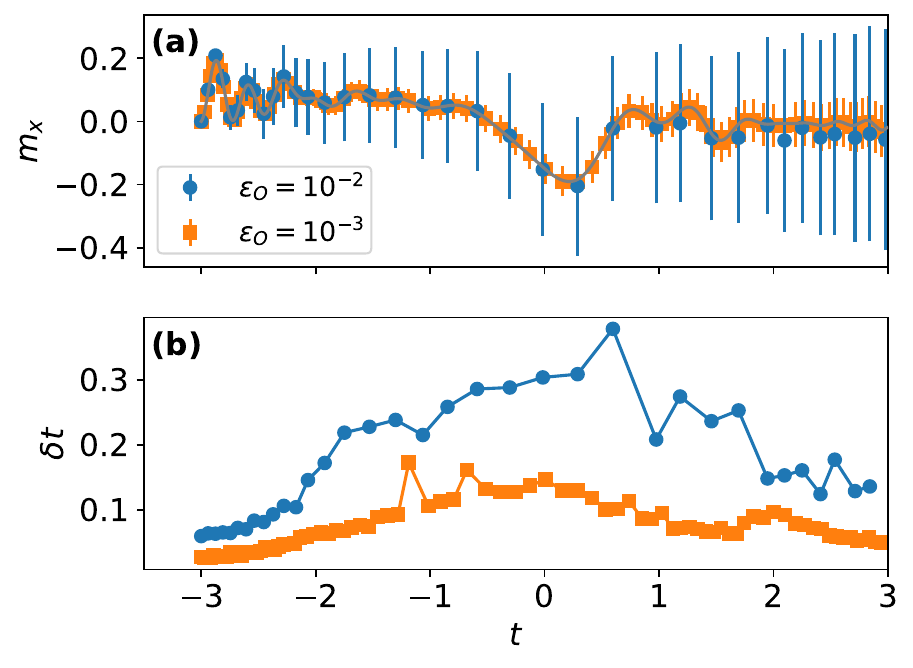}
    \caption{(a) Dynamics of $x$-magnetization density, $m_x$, under the time-dependent Hamiltonian~\eqref{eq:Htdep}, calculated by observable-based Trotter24 for $O=m_x$ and tolerance $\epsilon_O=10^{-2}$ (circle) and $10^{-3}$ (square). The error bar shows the theoretical upper bound~\eqref{eq:boundO2}. The solid curve shows an accurate solution obtained by the fourth-order Trotterization with $\dt=10^{-2}$, the system size is $L=18$, and the safety constant is $C=0.95$.
    (b) Stepsize $\dt$ chosen adaptively in the algorithm in each step. Different symbols correspond to those in panel (a).}
    \label{fig:demo_tdep}
\end{figure}

Let us now implement Trotter24 in an example time-dependent Hamiltonian.
Our model is a generalization of Eq.~\eqref{eq:Hpm} as
\begin{align}\label{eq:Htdep}
    H(t) = t A + B,
\end{align}
where $A$ and $B$ are given in Eq.~\eqref{eq:Hpm}.
We take the same initial state as in Sec.~\ref{sec:benchmark} and set the time interval as $t_\mathrm{ini}=-3.0$ and $t_\mathrm{fin}=+3.0$.
Here we only demonstrate the observable-based one since the fidelity-based one works similarly.
Figure~\ref{fig:demo_tdep} shows the $x$-magnetization density dynamics obtained by Trotter24 for different tolerance $\epsilon_O=10^{-2}$ and $10^{-3}$.
Like in the time-independent case, Trotter24 provides the dynamics in a precision-guaranteed way, in which the exact solution resides in the error bars given by the preset tolerance $\epsilon_O$.

Unlike in the time-independent case, the stepsize tends to decrease as $|t|$ increases, as shown in Fig.~\ref{fig:demo_tdep}(b).
This is consistent with the fact that the Hamiltonian~\eqref{eq:Htdep}, or the energy scale, is proportional to $t$ when $|t|\gg1$, for which the stepsize needs to be decreased to keep the error below the tolerance.
Trotter24 automatically chooses the appropriate stepsize, depending on the instantaneous Hamiltonian as well as the instantaneous quantum state.

\section{Conclusions and Discussions}
We have developed a method of measuring the Trotter error by combining it with another higher-order Trotterization without ancillary qubits.
Using this, we devised an algorithm named Trotter$(m,n)$ for Hamiltonian simulations, during which the stepsize $\dt$ is adaptively chosen as large as possible within an error tolerance we set in advance.
In each time step, the precision is guaranteed in the sense of Eqs.~\eqref{eq:Fupper} and ~\eqref{eq:boundO2} with higher-order corrections being neglected.
Our algorithm applies to both time-independent and --dependent Hamiltonians, as we benchmarked in the example Hamiltonians in the quantum spin chain mainly for Trotter24, i.e., the case of $(m,n)=(2,4)$.
Another merit of Trotter24 is the efficiency in finding the optimal stepsize $\dt$.
According to the benchmark, the estimated $\dt$ is rejected less than once on average during each time step by setting the safety constant $C\le 0.99$.
In exchange for conducting measurements in each step, Trotter24 adaptively finds almost the largest $\dt$, which was about ten times as large as that inferred from the upper bound arguments. 
Thus, this algorithm keeps the circuit significantly shallower within our error tolerance $\epsilon$.

{\hl
Although we focused on applying Trotter$(m,n)$ to quantum computers, this algorithm also applies to classical computations.
Beyond the system sizes accessible with the full exact diagonalization of Hamiltonians, Trotterization-based algorithms, such as the time-evolution block-decimation~\cite{Vidal2004} and the full state-vector evolution~\cite{Jones2019}, are useful on high-performance computers. When we use an $m$-th order Trotterization, one may want to estimate its error and guarantee that the error is within a tolerance $\epsilon$. For such a purpose, Trotter$(m,n)$ provides these classical computations with a precision-guaranteed adaptive stepsize method, enabling reliable simulations for quantum many-body dynamics, although there could be other possible methods.
}

We have neglected the statistical error in measuring the Trotter error through sampling and the device error inherent to NISQ computers.
These errors are, in principle, estimated from the available number of measurements and the device assessment.
Also, the error mitigation technique~\cite{Endo2018,Cai2022} helps to reduce these errors, as demonstrated on a 100-qubit-scale NISQ computer~\cite{Kim2023}.
If these errors are below our tolerance and the Trotter error is the bottleneck, Trotter24 will benefit us in optimizing Trotterization in a precision-guaranteed manner.
We leave real-device implementations for future work.

\section*{Acknowledgements}
Fruitful discussions with H. Zhao, J. Ostmeyer, and K. Mizuta are gratefully acknowledged.
T. N. I. was supported by JST PRESTO Grant No. JPMJPR2112 and by JSPS KAKENHI Grant No. JP21K13852.
A part of numerical calculations has been performed using Qulacs~\cite{Suzuki2021}.
This work is supported by MEXT
Quantum Leap Flagship Program (MEXTQLEAP)
Grant No. JPMXS0118067394, JPMXS0120319794, and
JST COI-NEXT program Grant No. JPMJPF2014.

\appendix
\section{Scalings of Trotter errors}\label{app:error}
\subsection{Leading-order expression for $\eta_F$}
According to the BCH formula, we have
\begin{align}\label{eq:UdagT2}
U(\dt)^\dag T_m(\dt)&= e^{i H\dt}e^{-i H\dt +\Upsilon_{m+1}}= e^{\tilde{\Upsilon}_{m+1}},
\end{align}
where $T_m(\dt)$ denotes an $m$-th order Trotterization and
\begin{align}\label{eq:Up3tilde}
    \tilde{\Upsilon}_{m+1} = \Upsilon_{m+1}+[iH\dt,\Upsilon_{m+1}]+\dots = \Upsilon_{m+1} + O(\dt^{m+2})
\end{align}
is an anti-Hermitian operator, and we used $\Upsilon_{m+1}=O(\dt^{m+1})$.
Thus, we have
\begin{widetext}
\begin{align}
    |\braket{\psi(t+\dt)|\psi_m(t+\dt)}|^2
    &=|\braket{\psi(t)|e^{\tilde{\Upsilon}_{m+1}}|\psi(t)}|^2\\
    &=|\braket{\psi(t)|\left[1+\tilde{\Upsilon}_{m+1}+\frac{1}{2}\tilde{\Upsilon}_{m+1}^2+O(\dt^{3(m+1)})\right]|\psi(t)}|^2\\
    &=1+\braket{\psi(t)|\tilde{\Upsilon}_{m+1}^2|\psi(t)}+|\braket{\psi(t)|\tilde{\Upsilon}_{m+1}|\psi(t)}|^2+O(\dt^{3(m+1)})\\
    &=1-\braket{\psi(t)|(i\tilde{\Upsilon}_{m+1})^2|\psi(t)}+\braket{\psi(t)|(i\tilde{\Upsilon}_{m+1})|\psi(t)}^2+O(\dt^{3(m+1)})\\
    &=1-\braket{\psi(t)|(i\Upsilon_{m+1})^2|\psi(t)}+\braket{\psi(t)|(i\Upsilon_{m+1})|\psi(t)}^2+O(\dt^{2m+3}),
\end{align}
\end{widetext}
where we used $\mathrm{Re}\braket{\psi(t)|\tilde{\Upsilon}_{m+1}|\psi(t)}=0$ (since $\tilde{\Upsilon}_{m+1}$ is anti-Hermitian) and Eq.~\eqref{eq:Up3tilde}.
Therefore, we obtain
\begin{align}
    \eta_F &= \sqrt{\braket{\psi(t)|(i\Upsilon_{m+1})^2|\psi(t)}-\braket{\psi(t)|(i\Upsilon_{m+1})|\psi(t)}^2}\notag\\
    &\qquad+O(\dt^{m+2}).
\end{align}
Notice that the right-hand side is nonnegative since $i\Upsilon_{m+1}$ is Hermitian.

\subsection{Fidelity difference between the exact and measurable expressions}
Here we study the difference between
\begin{align}
    \chi &\equiv \braket{\psi(t+\dt)|\psi_m(t+dt)},\\
    \chi_{mn} &\equiv \braket{\psi_n(t+\dt)|\psi_m(t+dt)}
\end{align}
and show $\eta_F=\eta_F^{(mn)}+O(\dt^{n+1})$.
We begin by noting
\begin{align}
    \chi-\chi_{mn} = \braket{\psi(t)| U^\dag(\dt)[1-U(\dt)T_n^\dag(\dt)]T_m(\dt)|\psi(t)}.
\end{align}
According to the BCH formula, we have
\begin{align}\label{eq:UdagT4}
U(\dt)^\dag T_n(\dt)&= e^{i H\dt}e^{-i H\dt +\Upsilon_{n+1}}= e^{\tilde{\Upsilon}_{n+1}},
\end{align}
where
\begin{align}\label{eq:Up5tilde}
    \tilde{\Upsilon}_{n+1} = \Upsilon_{n+1} + [iH\dt, \Upsilon_{n+1}]+\cdots =O(\dt^{n+1}) 
\end{align}
is an anti-Hermitian operator, and we used $\Upsilon_{n+1}=O(\dt^{n+1})$.
Thus we have
\begin{align}
    \chi-\chi_{mn} &= \braket{\psi(t)| U^\dag(\dt)\tilde{\Upsilon}_{n+1} T_{m}(\dt)|\psi(t)}+O(\dt^{2(n+1)})\\
&= \braket{\psi(t)| T_m^\dag(\dt)\tilde{\Upsilon}_{n+1} T_m(\dt)|\psi(t)}+O(\dt^{m+n+2}),
\end{align}
where we used $U(\dt)=T_m(\dt)+O(\dt^{m+1})$ and $\tilde{\Upsilon}_{n+1}=O(\dt^{n+1})$.
Since $\tilde{\Upsilon}_{n+1}$ is anti-Hermitian, 
\begin{align}
    \delta \chi \equiv \braket{\psi(t)| T_m^\dag(\dt)\tilde{\Upsilon}_{n+1} T_m(\dt)|\psi(t)}=O(\dt^{n+1})
\end{align}
is pure imaginary.

Now we rewrite $\eta_F$ using $\chi_{mn}$ as follows,
\begin{align}
    \eta_F^2 &= 1-|\chi|^2 = 1-|\chi_{mn}+\delta \chi|^2\\
    &= 1-|\chi_{mn}|^2 -(\chi_{mn}^* \delta \chi + \mathrm{c.c.})+O(\dt^{2(n+1)}).
\end{align}
Here we note 
\begin{align}
    \chi_{mn} &= \braket{\psi(t)| T_n(\dt)^\dag T_m(\dt) |\psi(t)}\\
    &= \braket{\psi(t)| [T_n(\dt)^\dag U(\dt)] [U^\dag(\dt)T_m(\dt)] |\psi(t)}\\
    &=\braket{\psi(t)| e^{-\tilde{\Upsilon}_{n+1}} e^{\tilde{\Upsilon}_{m+1}} |\psi(t)}\label{eq:chi24-a}\\
    &=1 +\braket{\psi(t)|\tilde{\Upsilon}_{m+1}|\psi(t)} + O(\dt^{n+1})\label{eq:chi24-b}
\end{align}
where we used Eqs.~\eqref{eq:UdagT2} and \eqref{eq:UdagT4} to have Eq.~\eqref{eq:chi24-a}, and Eq.~\eqref{eq:Up5tilde} to obtain Eq.~\eqref{eq:chi24-b}.
Substituting Eq.~\eqref{eq:chi24-b} and using the facts $\mathrm{Re}\delta \chi=0$, $\delta\chi=O(\dt^{n+1})$, and $\tilde{\Upsilon}_{m+1}=O(\dt^{m+1})$, we obtain
\begin{align}
    \eta_F^2 &= 1 - |\chi_{mn}|^2 +O(\dt^{m+n+2}),\\
    \eta_F &= \eta^{(mn)}_F + O(\dt^{n+1}),
\end{align}
where we used $\eta^{(mn)}_F = O(\dt^{m+1})$.

\subsection{Observable difference between the exact and fourth-order expressions}
Here we show 
\begin{align}
    \eta_O - \eta_O^{(mn)} = O(\dt^{n+1}).
\end{align}
This is simply obtained from
\begin{align}
&\braket{\psi_n(t+\dt)|O|\psi_n(t+\dt)} \notag\\
&=\braket{\psi(t)|e^{iH\dt -\Upsilon_{n+1}}Oe^{-iH\dt +\Upsilon_{n+1}}|\psi(t)}\\
&=\braket{\psi(t)|e^{iH\dt}e^{-\Upsilon_{n+1}}Oe^{\Upsilon_{n+1}}e^{-iH\dt}|\psi(t)}+O(\dt^{n+2})\\
&=\braket{\psi(t+\dt)|O|\psi(t+\dt)}-\braket{\psi(t)|[\Upsilon_{n+1},O]|\psi(t)}+O(\dt^{n+2})\\
&=\braket{\psi(t+\dt)|O|\psi(t+\dt)}+O(\dt^{n+1}),
\end{align}
which means
\begin{align}
&\eta_O - \eta_O^{(24)}\notag\\
&=\braket{\psi(t+\dt)|O|\psi(t+\dt)}-\braket{\psi_4(t+\dt)|O|\psi_4(t+\dt)}\notag\\
&=O(\dt^{n+1}).
\end{align}

\section{Fidelity Error Propagation}\label{app:error_propagation}
To prove Eq.~\eqref{eq:Fupper} for a general order $m$ (we obtain a proof of Eq.~\eqref{eq:Fupper} by setting $m=2$ in the following argument), we begin by introducing 
\begin{align}
    \delta U(\dt) \equiv T_m^\dag(\dt)U(\dt)-1=e^{-\tilde{\Upsilon}_{m+1}(\dt)}-1 = O(\dt^{m+1}),
\end{align}
where we used Eq.~\eqref{eq:Up3tilde} and explicitly showed the $\dt$-dependence of $\tilde{\Upsilon}_{m+1}$.
Introducing 
\begin{align}
    \Gamma_{m+1}(\dt) \equiv i\tilde{\Upsilon}_{m+1}(\dt)=O(\dt^{m+1}),
\end{align}
which is Hermitian, we have
\begin{align}
    \delta U(\dt) &= e^{i\Gamma_{m+1}(\dt)}-1\\
    &= i\Gamma_{m+1}(\dt) -\frac{\Gamma_{m+1}(\dt)^2}{2}+O(\dt^{3(m+1)}).
\end{align}

Now we define
\begin{align}
    \chi_N \equiv \braket{\psi(t_N)|\psi_2(t_N)},
\end{align}
which satisfies
\begin{align}
    \delta F_N \equiv 1-|\chi_N|^2 = \eta_{F,N}^2.
\end{align}
Then we have
\begin{widetext}
\begin{align}
    \chi_N &= \bra{\psi_0} \prod_{i=0,\dots,N-1}^{\rightarrow}\left\{T_m^\dag(\dt)\left[1+i\Gamma_{m+1}(\dt_i)-\frac{\Gamma_{m+1}(\dt_i)^2}{2}\right]\right\}
    \prod_{i=0,\dots,N-1}^{\leftarrow}T_m(\dt_i)\ket{\psi_0}\\
    &=1-i\sum_{i=1}^{N}\braket{\psi(t_i)|\Gamma_{m+1}(\dt_i)|\psi(t_i)}-\frac{1}{2}\sum_{i=1}^{N}\braket{\psi(t_i)|\Gamma_{m+1}(\dt_i)^2|\psi(t_i)}\notag\\
    &\qquad\qquad -\sum_{1\le i<j\le N}\braket{\psi(t_i)|\Gamma_{m+1}(\dt_i)T_m^\dag(\dt_{i+1})\cdots T_m^\dag(\dt_{j})\Gamma_{m+1}(\dt_j) |\psi(t_j)}+O(\dt^{3(m+1)});\\
    \delta F_N &= \sum_{i=1}^{N}\braket{\psi(t_i)|\Gamma_{m+1}(\dt_i)^2|\psi(t_i)}
    +2\mathrm{Re}\left[ \sum_{1\le i<j\le N}\braket{\psi(t_i)|\Gamma_{m+1}(\dt_i)T_m^\dag(\dt_{i+1})\cdots T_m^\dag(\dt_{j})\Gamma_{m+1}(\dt_j) |\psi(t_j)} \right]\notag\\
    &\qquad\qquad -\left(\sum_{i=1}^{N}\braket{\psi(t_i)|\Gamma_{m+1}(\dt_i)|\psi(t_i)} \right)^2+O(\dt^{3(m+1)}),
\end{align}
\end{widetext}
where we used $\mathrm{Im}\braket{\psi(t_i)|\Gamma_{m+1}(\dt_i)|\psi(t_i)}=0$.
Here we notice that
\begin{widetext}
\begin{align}
&\| T_m(\dt_j)\cdots T_m(\dt_i)\Gamma_{m+1}(\dt_i)\ket{\psi(t_i)}-\Gamma_{m+1}(\dt_j)\ket{\psi(t_j)}\|^2\notag\\
&\qquad=\braket{\psi(t_i)|\Gamma_{m+1}(\dt_i)^2|\psi(t_i)}+\braket{\psi(t_j)|\Gamma_{m+1}(\dt_j)^2|\psi(t_j)}\notag\\
&\qquad\qquad -2\mathrm{Re}[\braket{\psi(t_i)|\Gamma_{m+1}(\dt_i)T_m^\dag(\dt_{i+1})\cdots T_m^\dag(\dt_{j})\Gamma_{m+1}(\dt_j) |\psi(t_j)}]
\end{align}
\end{widetext}
is higher-order and negligible in our leading-order calculations, meaning that we can replace $2\mathrm{Re}[\braket{\psi(t_i)|\Gamma_{m+1}(\dt_i)T_m^\dag(\dt_{i+1})\cdots T_m^\dag(\dt_{j})\Gamma_{m+1}(\dt_j) |\psi(t_j)}]$ by $\braket{\psi(t_i)|\Gamma_{m+1}(\dt_i)^2|\psi(t_i)}+\braket{\psi(t_j)|\Gamma_{m+1}(\dt_j)^2|\psi(t_j)}$ neglecting higher-order corrections.
By doing so and completing squares, we obtain
\begin{widetext}
\begin{align}
    \delta F_N &\approx N\sum_{i=1}^N [\braket{\psi(t_i)|\Gamma_{m+1}(\dt_i)^2|\psi(t_i)} - \braket{\psi(t_i)|\Gamma_{m+1}(\dt_i)|\psi(t_i)}^2]\notag\\
    &\qquad +\sum_{1\le i<j\le N} (\braket{\psi(t_i)|\Gamma_{m+1}(\dt_i)|\psi(t_i)}-\braket{\psi(t_j)|\Gamma_{m+1}(\dt_j)|\psi(t_j)})^2.\label{eq:dF_app}
\end{align}
\end{widetext}
We notice again that the second term on the right-hand side of Eq.~\eqref{eq:dF_app} is negligible in the leading-order calculation.
Recalling Eq.~\eqref{eq:eta_F_leading}, we notice that the first term consists of the leading-order fidelity error in each time step, which is guaranteed to be less than $\epsilon^2$ in the algorithm.
Therefore we finally obtain
\begin{align}
    \delta F_N &\lesssim N^2 \epsilon^2,\\
    \eta_{F,N} &\lesssim N\epsilon.
\end{align}

\section{Error-bound approach}\label{sec:errorbound}
Here we apply the exact error bound~\cite{Kivlichan2020} for our example model, obtaining $\dt$ guaranteeing the error is less than our tolerance $\epsilon$.
According to Ref.~\cite{Kivlichan2020}, we have the following inequality
\begin{align}\label{eq:boundAB}
    \| U(\dt) - e^{-i A \dt/2}e^{-i B \dt}e^{-i A \dt/2}\| \le W_{A,B} \dt^3,
\end{align}
where
\begin{align}
    W_{A,B} \equiv \| [B,[B,A]]\| + \frac{1}{2}\| [A,[B,A]]\|.
\end{align}
Note that interchanging $A$ and $B$ also leads to
\begin{align}\label{eq:boundBA}
    \| U(\dt) - e^{-i B \dt/2}e^{-i A \dt}e^{-i B \dt/2}\| \le W_{B,A} \dt^3.
\end{align}
If $W_{A,B}<W_{B,A}$, Inequality~\eqref{eq:boundAB} gives a tighter bound, and one may use the second-order formula $e^{-i A \dt/2}e^{-i B \dt}e^{-i A \dt/2}$ in this order of $A$ and $B$.
Otherwise, one may use $e^{-i B \dt/2}e^{-i A \dt}e^{-i B \dt/2}$ whose error is bounded by Inequality~\eqref{eq:boundBA}

\begin{figure}
    \centering
    \includegraphics[width=\columnwidth]{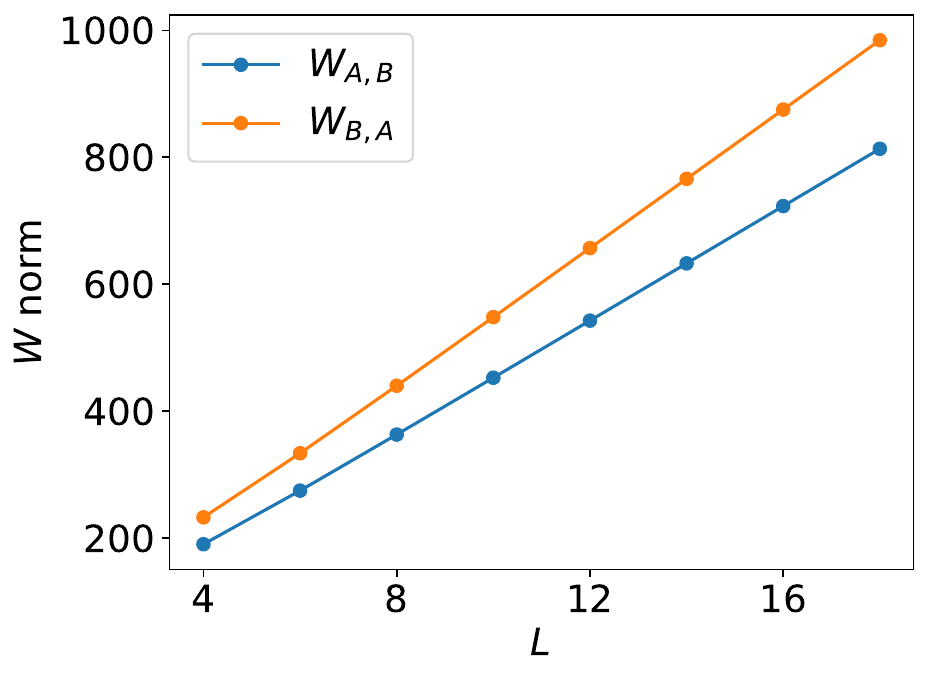}
    \caption{System-size dependence of the $W$ norms for the example model~\eqref{eq:Hpm}}
    \label{fig:Wnorm}
\end{figure}

The $W$ norms, $W_{A,B}$ and $W_{B,A}$, for our example model~\eqref{eq:Hpm} discussed in the main text are plotted in Fig.~\ref{fig:Wnorm}.
Recall that we focused on $e^{-i A \dt/2}e^{-i B \dt}e^{-i A \dt/2}$ in the main text, and this is consistent with the tighter bound~\eqref{eq:boundAB} since $W_{A,B}<W_{B,A}$.

The error-bound approach based on the bound~\eqref{eq:boundAB} determines $\dt$ from
\begin{align}
    W_{A,B}\dt^3 \le \epsilon.
\end{align}
The possible maximum for $\dt$ satisfying this inequality is denoted by $\dt_\mathrm{bound}$ and given in Eq.~\eqref{eq:dt_bound}.

\section{Comparison with Richardson's extrapolation}\label{sec:extrapolation}
In this appendix, we compare Trotter24 and Richardson's extrapolation as follows.
We first implement the Trotter24 like in Sec.~\ref{sec:benchmark} to obtain the magnetization expectation values at times $t_N$ ($N=1,2,\dots$).
On the other hand, for each time $t_N$, we aim to obtain a good estimate for the exact expectation value using Richardson's extrapolation.
Namely, for each integer $M$ taken out of $m+1$ integers, we implement the usual second-order Trotterization with stepsize $t_N/M$, having expectation values at $t_N$.
To equate the maximum required gate complexity in both methods, we impose $M\le N$.

\begin{figure}
    \centering
    \includegraphics[width=\columnwidth]{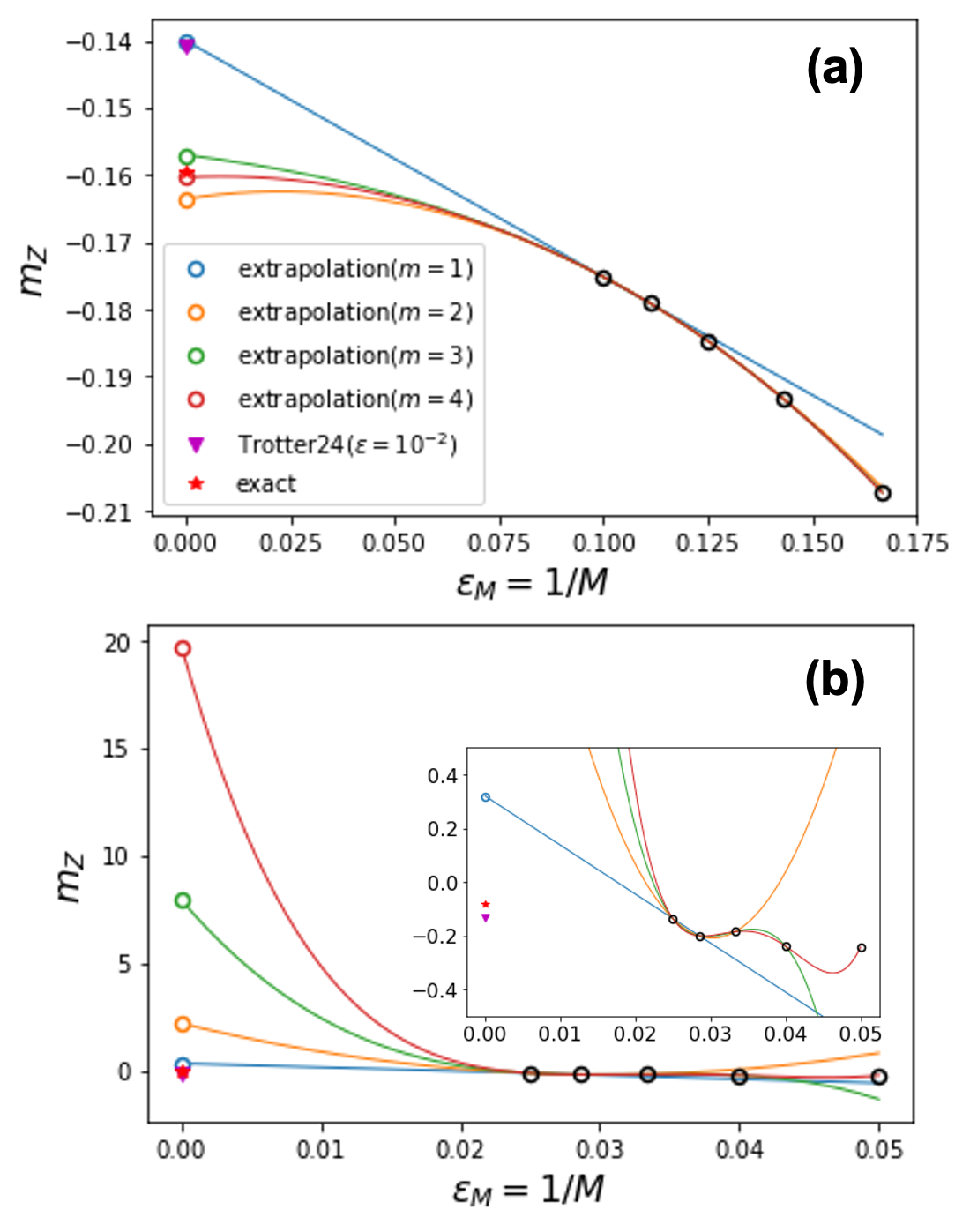}
    \caption{Comparison between Trotter24 and Richardson's extrapolation at $t_{10}=1.68$ (a) and $t_{40}=10.3$ (b).
    Filled symbols show the magnetization expectation values obtained by the exact calculation (star) and by the observable-based Trotter24 with $\varepsilon=10^{-2}$.
    Open black circles show those obtained by the second-order Trotterization, $m$ of which are polynomially extrapolated to estimate the ideal limit $\varepsilon_M\to0$ (colored circles shown in the legends).
    }
    \label{fig:time_dependence}
\end{figure}

Figure~\ref{fig:time_dependence} shows the comparison at $t=t_{10}$ and $t_{40}$, where Trotter24 is observable-based with $\varepsilon=10^{-2}$.
Panel (a) shows results for the shorter $t_{10}=1.68$.
While the accuracy for $m=1$ is comparable to Trotter24, the extrapolation gives a better estimation as $m$ increases to outperform Trotter24.
In contrast, Panel (b) shows the results for the longer $t_{40}=10.3$, where the extrapolation gives a worse estimation as $m$ increases.

This breakdown of extrapolations for longer times is explained by Runge's phenomenon.
The expectation values of $O$ obtained by the second-order Trotterization reads
\begin{equation}
    O(t,\varepsilon_{M})= \Braket{\psi(0)| T_{2}^\dag(\varepsilon_M t)^M O T_{2}(\varepsilon_M t)^{M} |\psi(0)},
\end{equation}
where $\varepsilon_M\equiv 1/M$, and we assume that $O(t,\varepsilon_M)$ is $(m+1)$-times differentiable with respect to $\varepsilon_M$.
Note that the exact value is given by
$O(t,0)=\Braket{\psi(0)|U(t)^{\dag}OU(t)|\psi(0)}=\lim_{M \rightarrow \infty}O(t,\varepsilon_M)$
The extrapolation method estimates this by extrapolating an $m$-th order polynomial curve going through the $(m+1)$ points, $(\varepsilon_{M_{0}},O(t,\varepsilon_{M_{0}})), \cdots, (\varepsilon_{M_{m}},O(t,\varepsilon_{M_{m}}))$, where we assume $M_{0} > \cdots > M_{m}$.
Thus, utilizing the coefficients $c_i$ obtained through Neville's algorithm, we can write the estimate as $\widetilde{O}_{m}(t,0)=\sum_{i=0}^{m}c_{i}O(t,\varepsilon_{M_{i}})$.
Then, the error of the estimate is bounded as
\begin{widetext}
\begin{equation}
    |O(t,0) - \widetilde{O}_{m}(t,0)| \leq \max_{0\leq \xi \leq \varepsilon_{M_{m}}} \frac{1}{(m+1)!} \left| \frac{\partial^{(m+1)} O(t,\xi)}{\partial \varepsilon_{M}^{(m+1)}} \right| \prod_{i=0}^{m}\varepsilon_{M_{i}}.
\end{equation}
\end{widetext}
Notice that 
$\max_{0\leq \xi \leq \varepsilon_{M_{m}}} \left| \frac{\partial^{(m+1)}}{\partial \varepsilon_{M}^{(m+1)}} O(t,\xi) \right| $ can increase when $m$ increases.
Consequently, the extrapolation estimates can be worse even though we increase $m$, and this is known as Runge's phenomenon.
This phenomenon can occur when $m$ is too large even if $t$ is short.
Generally speaking, $O(t,\varepsilon_M)$ tends to become a more complex function of $\varepsilon_M$ as $t$ increases, leading to instability.

Runge's phenomenon is circumvented, in classical numerics, by optimizing the sequence $\varepsilon_0,\varepsilon_1,\dots,\varepsilon_m$ such as the Chebyshev notes. 
However, this type of optimization is nontrivial in NISQ devices because of the limitation of circuit depth and the constraint that $M_i$'s are integers.
In fact, a recent study~\cite{rendon2022improved} resorts to a beyond-NISQ quantum computation for solving the optimization.
Trotter24 is free from Runge's phenomenon and more stable especially in longer times.

\end{document}